\documentclass[journal,10pt,twocolumn,letter]{IEEEtran}

\usepackage{graphics}
\usepackage{colortbl}
\usepackage{multicol}
\usepackage{lipsum}
\usepackage{color}
\usepackage[cmex10]{amsmath}
\usepackage{amsthm}
\usepackage{amssymb}
\usepackage{mathrsfs}
\usepackage{mathtools}
\usepackage{amsbsy}
\usepackage[colorlinks=true,bookmarks=false,citecolor=blue,urlcolor=blue]{hyperref}
\usepackage{xcolor}
\usepackage{mathrsfs}
\usepackage{setspace}
\usepackage{float}
\usepackage{graphicx}
\usepackage{pstool}
\usepackage{lettrine}
\usepackage{newclude}
\usepackage[normalem]{ulem}
\usepackage{latexsym}
\usepackage{algpseudocode}
\usepackage{gensymb}

\usepackage{balance}

\usepackage{multirow}
\usepackage{amsmath}
\usepackage{amsmath,bm}
\usepackage{wasysym}
\usepackage{mathrsfs}
\usepackage{amsmath,cite,amsfonts,amssymb,color}

\usepackage[linesnumbered,ruled,vlined]{algorithm2e}
\SetKwInput{KwInput}{Input}  
\SetKwInput{KwOutput}{Output}
\SetKw{KwBy}{by}
\makeatletter
\newcommand{\nosemic}{\renewcommand{\@endalgocfline}{\relax}}
\newcommand{\dosemic}{\renewcommand{\@endalgocfline}{\algocf@endline}}
\let\oldnl\nl
\newcommand{\nonl}{\renewcommand{\nl}{\let\nl\oldnl}}
\makeatother
\ifCLASSOPTIONcompsoc
\usepackage[caption=false,font=normalsize,labelfon
t=sf,textfont=sf]{subfig}
\else
\usepackage[caption=false,font=footnotesize]{subfig}
\fi

\allowdisplaybreaks

\newtheorem{proposition}{Proposition}

\newenvironment{skproof}{%
  \proof}{\endproof}
\graphicspath{{figures/}}
\allowdisplaybreaks

\newcommand{\RNum}[1]{\uppercase\expandafter{\romannumeral #1\relax}}

\usepackage{mathtools}

\usepackage{accents}
\newlength{\dhatheight}
\newlength{\dtildeheight}
\newcommand{\doublehat}[1]{%
    \settoheight{\dhatheight}{\ensuremath{\hat{#1}}}%
    \addtolength{\dhatheight}{-0.2ex}%
    \hat{\vphantom{\rule{1pt}{\dhatheight}}%
    \smash{\hat{#1}}}}
\newcommand{\doubletilde}[1]{%
   \settoheight{\dtildeheight}{\ensuremath{\hat{#1}}}%
   \addtolength{\dtildeheight}{-0.15ex}%
   \tilde{\vphantom{\rule{1pt}{\dtildeheight}}%
   \smash{\tilde{#1}}}}   

\begin{document}

\title{UAV-Assisted Underwater Sensor Networks\\using RF and Optical Wireless Links}

\author{\IEEEauthorblockN{Pouya Agheli\thanks{P. Agheli, H. Beyranvand, and M. J. Emadi are with the Department of Electrical Engineering, Amirkabir University of Technology (Tehran Polytechnic), Tehran, Iran (E-mails: \{pouya.agheli, beyranvand, mj.emadi\}@aut.ac.ir).},
Hamzeh Beyranvand,
and Mohammad Javad Emadi
}} 


\maketitle

\begin{abstract}
Underwater sensor networks (UWSNs) are of interest to gather data from underwater sensor nodes (SNs) and deliver information to a terrestrial access point (AP) in the uplink transmission, and transfer data from the AP to the SNs in the downlink transmission. In this paper, we investigate a triple-hop UWSN in which autonomous underwater vehicle (AUV) and unmanned aerial vehicle (UAV) relays enable end-to-end communications between the SNs and the AP. It is assumed that the SN--AUV, AUV--UAV, and UAV--AP links are deployed by underwater optical communication (UWOC), free-space optic (FSO), and radio frequency (RF) technologies, respectively. Two scenarios are proposed for the FSO uplink and downlink transmissions between the AUV and the UAV, subject to water-to-air and air-to-water interface impacts; direct transmission scenario (DTS) and retro-reflection scenario (RRS). After providing the channel models and their statistics, the UWSN's outage probability and average bit error rate (BER) are computed. Besides, a tracking procedure is proposed to set up reliable and stable AUV--UAV FSO communications. Through numerical results, it is concluded that the RSS scheme outperforms the DTS one with about $200\%$ ($32\%$) and $80\%$ ($17\%$) better outage probability (average BER) in the uplink and downlink, respectively. It is also shown that the tracking procedure provides on average $480\%$ and $170\%$ improvements in the network's outage probability and average BER, respectively, compared to poorly aligned FSO conditions. The results are verified by applying Monte-Carlo simulations.
\end{abstract}

\IEEEpeerreviewmaketitle

\section{Introduction}
\lettrine{U}{nderwater} sensor networks (UWSNs) enable biological observations, safe navigation, the study of subaquatic animals and plants, and oil spills' positioning. The goal is to set up a reliable and fast delivery of sensing data via underwater sensor nodes (SNs) to a terrestrial center in the uplink, as well as the command data from the center to the SNs in the downlink with the lowest outage probability and average bit error rate (BER).  The underwater optical communication (UWOC) is a promising technology for collecting data from the SNs distributed at the bottom of a sea, while the radio frequency (RF) and acoustic carriers suffer from high latency and low data rates \cite{christopoulou2019outage,kaushal2016underwater,amer2019underwater}. Besides, the under- and above-water relays can support the line-of-sight (LOS) transmission requirement of optical wireless links, provide long-distance communications, tackle the high absorption and scattering of optical signals in the water, and minimize the transmission power at the SNs \cite{christopoulou2019outage} and \cite{vavoulas2014underwater}. To this end, different system models have been proposed to obtain reliable wireless--optical connections between the SNs and the terrestrial center \cite{christopoulou2019outage,amer2019underwater}, and \cite{naik2020co,anees2019performance,lei2020performance,aneeshybrid,li2019wdm,jurado2019converging}. Specifically, \cite{amer2019underwater} and  \cite{anees2019performance,naik2020co,lei2020performance} have suggested dual-hop networks in which RF and UWOC links connect buoyant relays to a terrestrial access point (AP) and underwater nodes, respectively. However, \cite{christopoulou2019outage} and \cite{aneeshybrid,li2019wdm,jurado2019converging} have used free-space optic (FSO) links to provide robust and low latency communications between buoyant relays and an AP.

The FSO technology offers high data rates with rapid setup time, easy upgrade, flexibility, freedom from spectrum license regulations, protocol transparency, and enhanced security \cite{abdalla2020optical,zedini2015multihop,tang2014multihop}. However, it comes at the expense of some drawbacks such as pointing error, the requirement of a LOS connection between the communicating nodes, and sensitivity to the atmospheric conditions such as rain, snow, fog, and dust \cite{khalighi2014survey,kaushal2016optical,agheli2021design}. To compensate for the outage issue of the FSO links in the adverse atmospheric conditions, the hybrid RF/FSO solution is introduced \cite{douik2016hybrid,touati2016effects,chen2016multiuser,usman2014practical,zhang2009soft,jamali2016link,najafi2017optimal}. Furthermore, in \cite{hassan2017statistical}, buffer-aided RF/FSO links have been utilized to enhance the network's performance in the unfavorable atmospheric conditions at the cost of increasing the delay. In \cite{agheli2021cognitive}, a cognitive RF--FSO fronthaul assignment algorithm is proposed to tackle FSO misalignment and unfavorable weather conditions. The performance of wireless networks based on FSO links has been investigated in \cite{morra2017impact} and \cite{morra2018mixed}, where the impacts of wireless co-channel interference and FSO pointing error have been taken into account. 

Low-cost and highly mobile unmanned aerial vehicles (UAVs) have been used for many diverse applications, e.g., disaster management, environmental monitoring, and cellular (or satellite) networks \cite{ye2018secure,andre2014application,hu2018unmanned}. Thanks to their structures, UAVs enable fast deployment, flexible reconfiguration, and LOS connections without complex infrastructure requirements \cite{zeng2016wireless} and \cite{mozaffari2016unmanned}. Furthermore, UAVs have been utilized for remote sensing and relaying systems, which gather data from multiple sensors via ground-to-air links and, in return, deliver command data over air-to-ground links \cite{andre2014application}. Likewise, \cite{lei2020performance} has analyzed a dual-hop RF--UWOC communication system in which a buoyant node relays data between a UAV and a submarine over RF and UWOC links, respectively.

For FSO and UWOC use cases, pointing error is a barrier to have highly reliable communications in harsh environments or with mobile transceivers, such as UAVs \cite{kaymak2018survey}. However, various pointing, acquisition, and tracking mechanisms have been proposed that maintain stable LOS connections for FSO mobile applications \cite{kaymak2018survey}. Specifically, one promising solution is to take advantage of a modulating retro-reflection (MRR) system that can be exerted for widespread applications, e.g., satellite, marine, and submarine communication networks \cite{yang2018performance,yang2017channel,li2017probability,kaymak2018survey}. In general, the MRR system is assembled with an optical modulator and a passive retro-reflector of which corner cube reflector (CCR) and cat's eye reflector (CER) are two frequently-used types \cite{kaymak2018survey}. According to \cite{yang2018performance} and \cite{yang2017channel}, single- and double-path MRR-assisted FSO fading channels have been modeled by using log-normal and Gamma-Gamma distributions for weak and moderate-to-strong turbulence levels, respectively. Furthermore, \cite{li2017probability} has studied the impacts of atmospheric and distance parameters on the performance of MRR-assisted FSO links.

The previous studies have analyzed double-hop UWSNs with buoyant relays, e.g., ships, and UWOC links that cannot reliably communicate with deeply-located sensors at the bottom of the sea due to the high absorption and scattering phenomena. Also, the RF or FSO links communicating with the terrestrial center are probably affected by obstacles, e.g., nearby ships, between the buoyant relays and the terrestrial center. To solve the aforementioned issues, we investigate a triple-hop network wherein an autonomous underwater vehicle (AUV) relay is connected to the deeply-located sensors by relatively shorter UWOC links, and a UAV relays data between the AUV and the terrestrial AP through FSO and RF links, respectively, to provide blockage-free communications. Nevertheless, no tracking system has been proposed in the recent studies on UAV relaying over the sea. Since the FSO links connect two under- and above-water relays, we take into account the water-to-air (W2A) and air-to-water (A2W) impacts on the FSO links \cite{nabavi2019empirical,nabavi2019performance}. The main contributions of the paper are summarized as follows.
\begin{itemize}
    \item We study two full-duplex (FD) transmission strategies for the AUV--UAV FSO links; the first one uses two independent links for the uplink and downlink transmissions, while the other is based on the MRR system, where the uplink signals are transmitted to the UAV by reflecting and modulating the received downlink signals at the AUV.
    \item For reliable and stable AUV--UAV FSO communications with minimum pointing error, a tracking procedure at the UAV is proposed under an $n$-step acquisition-and-tracking algorithm with two tracking modes.
    \item Closed-form end-to-end outage probability and average BER expressions are derived for the uplink and downlink transmissions. To do so, we obtain the channel statistics, signal-to-noise ratio (SNR), outage probability, and average BER at each hop.
    \item Through numerical results, the network's performance is investigated from the outage probability and average BER perspectives, which are verified by using Monte-Carlo simulations. Also, the effects of various physical conditions and the AUV--UAV tracking procedure on the network's performance are studied.
\end{itemize}

\textit{Organization}: Section \ref{Sec:Sec2} introduces the channel models and their corresponding statistics at each hop of the UWSN in the presence of the AUV and UAV relays. The AUV--UAV tracking procedure and performance analyses are represented in Section \ref{Sec:Sec3}. Numerical results and discussions are presented in Section \ref{Sec:Sec4}. Finally, the paper is concluded in Section \ref{Sec:Sec5}.

\textit{Notation}: $[\,\cdot\,]^T$ stands for the transpose, and $(\,\cdot\,)^{-1}$ presents the inverse operator. Also, $\operatorname{erf}(x)\!=\!\frac{2}{\sqrt{\pi}}\int_{0}^{x}e^{-t^2}dt$ and $\operatorname{erfc}(x)\!=\!1\!-\operatorname{erf}(x)$ indicate the error and complementary error functions, respectively. In addition, $\operatorname{\gamma}(s,x)\!=\!\int_{0}^{x}t^{s-1} e^{-t} dt$ and $\Gamma(s,x)\!=\!\int_{x}^{\infty}t^{s-1} e^{-t} dt$ are the lower and upper incomplete Gamma functions, respectively, and $\operatorname{\Phi}(x)\!=\!\frac{1}{2}[1\!+\operatorname{erf}(\frac{x}{\sqrt{2}})]$ denotes the standard normal cumulative distribution function (CDF). Moreover, $\bm{x} \! \in \! \mathbb{C}^{n \times 1}$ denotes a vector in an $n$-dimensional complex space, $\mathbb{E}\{\cdot\}$ is the statistical expectation, $y \!\sim\! \mathcal{N}(m,\sigma^2)$ and $z \!\sim\! \mathcal{CN}(m,\sigma^2)$ respectively show real-valued and complex symmetric Gaussian random variables (RVs) with mean $m$ and variance $\sigma^2$.

\section{System Model} \label{Sec:Sec2}
\begin{figure}[t!]
    \centering
    \pstool[scale=0.55]{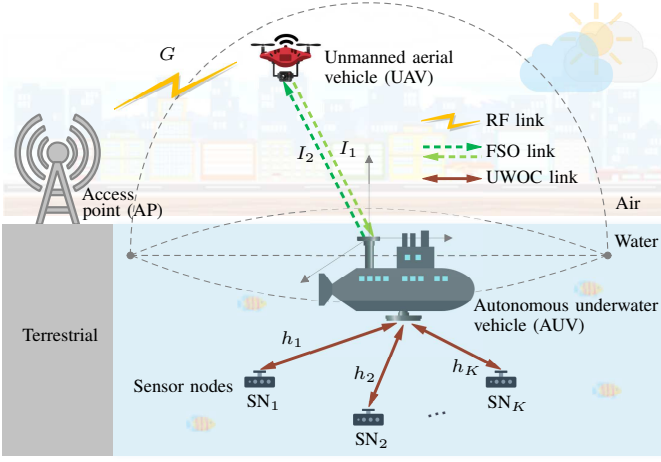}{
    \psfrag{G}{\vspace{-0.5cm}\hspace{0cm}\scriptsize $G$}
    \psfrag{Iu}{\hspace{0cm}\scriptsize $I_{2}$}
    \psfrag{Id}{\hspace{0cm}\scriptsize $I_{1}$}
    \psfrag{Air}{\hspace{0cm}\scriptsize Air}
    \psfrag{Water}{\hspace{0cm}\scriptsize Water}
    \psfrag{Sensor}{\scriptsize}
    \psfrag{Node}{\hspace{-0.4cm}\scriptsize Sensor nodes}
    \psfrag{h1}{\hspace{0cm}\scriptsize $h_1$}
    \psfrag{h2}{\hspace{0cm}\scriptsize $h_2$}
    \psfrag{hK}{\hspace{0cm}\scriptsize $h_K$}
    \psfrag{RFLink}{\hspace{0cm}\scriptsize RF link}
    \psfrag{FSOLink}{\hspace{0cm}\scriptsize FSO link}
    \psfrag{UWOCLink}{\hspace{0cm}\scriptsize UWOC link}
    \psfrag{SN1}{\hspace{0cm}\scriptsize $\text{SN}_1$}
    \psfrag{SN2}{\hspace{0cm}\scriptsize $\text{SN}_2$}
    \psfrag{SNK}{\hspace{0cm}\scriptsize $\text{SN}_K$}
    \psfrag{UAV1}{\hspace{-0.1cm}\scriptsize Unmanned aerial}
    \psfrag{UAV2}{\hspace{-0.1cm}\scriptsize vehicle (UAV)}
    \psfrag{AUV1}{\hspace{-0.2cm}\scriptsize Autonomous underwater}
    \psfrag{AUV2}{\hspace{-0.2cm}\scriptsize vehicle (AUV)}
    \psfrag{Terrestrial}{\hspace{-1.3cm}\scriptsize Terrestrial}
    \psfrag{Access}{\hspace{-0.08cm}\scriptsize Access}
    \psfrag{Point}{\hspace{-0.08cm}\scriptsize point (AP)}
    }
    \vspace{-0.25cm}
    \caption{The proposed UAV-assisted UWSN with RF and optical wireless links.}
    \label{Fig:Fig.1}
\end{figure}

We assume a triple-hop UWSN in which $K$ SNs are connected to an AUV through UWOC links, the AUV is connected to a UAV via FSO links, and the UAV is connected to a terrestrial wireless AP with an RF link, c.f. Fig.~\ref{Fig:Fig.1}. For the proposed system model, the following assumptions are made.
\begin{itemize}
    \item The UWOC and FSO links are deployed based on the wavelength-division multiplexing (WDM) technique.
    \item The uplink and downlink transmissions over the UWOC and RF links are established under the time-division duplexing (TDD) scheme.
    \item Two strategies are assumed for the FSO FD transmissions; direct transmission strategy (DTS) and retro-reflection strategy (RRS). In the DTS, uplink and downlink transmissions are performed over independent links. However, in the RRS, the uplink transmission is performed by reflecting the received downlink FSO beams on the UAV by an MRR terminal at the AUV.
    \item {On-off keying (OOK) modulation is applied for the uplink and downlink transmissions through all nodes.}
\end{itemize}

\begin{figure}[t!]
    \centering
    \pstool[scale=0.75]{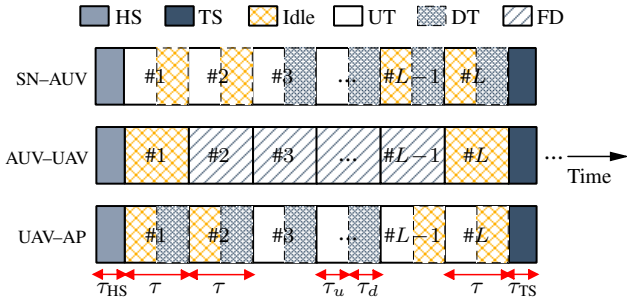}{
    \psfrag{SS}{\hspace{0cm}\scriptsize \textcolor{white}{HD}}
    \psfrag{ES}{\hspace{0cm}\scriptsize \textcolor{white}{TD}}
    \psfrag{A4}{\hspace{-0.04cm}\small $\tau_{\text{HS}}$}
    \psfrag{A1}{\hspace{0.04cm}\small $\tau$}
    \psfrag{A5}{\hspace{-0.04cm}\small $\tau_{\text{TS}}$}
    \psfrag{A2}{\hspace{0.0cm}\small $\tau_{u}$}
    \psfrag{A3}{\hspace{0.0cm}\small $\tau_{d}$}
    \psfrag{UT}{\hspace{0.04cm}\scriptsize UT}
    \psfrag{DT}{\hspace{0.04cm}\scriptsize DT}
    \psfrag{FDx}{\hspace{0.04cm}\scriptsize FD}
    \psfrag{...}{\hspace{-0.15cm} ...}
    \psfrag{-}{\hspace{-0.12cm} -}
    \psfrag{hop1}{\hspace{-0.11cm}\scriptsize SN--AUV}
    \psfrag{hop2}{\hspace{-0.25cm}\scriptsize AUV--UAV}
    \psfrag{hop3}{\hspace{-0.08cm}\scriptsize UAV--AP}
    \psfrag{Time}{\hspace{0.05cm}\footnotesize Time}
    \psfrag{HD}{\hspace{0.04cm}\footnotesize HS}
    \psfrag{TD}{\hspace{0.04cm}\footnotesize TS}
    \psfrag{Idle}{\hspace{0.04cm}\footnotesize Idle}
    \psfrag{UT}{\hspace{0.04cm}\footnotesize UT}
    \psfrag{DT}{\hspace{0.04cm}\footnotesize DT}
    \psfrag{FDx}{\hspace{0.04cm}\footnotesize FD}
    \psfrag{1}{\hspace{-0.21cm} \footnotesize {\#$1$}}
    \psfrag{2}{\hspace{-0.21cm} \footnotesize {\#$2$}}
    \psfrag{3}{\hspace{-0.21cm} \footnotesize {\#$3$}}
    \psfrag{L}{\hspace{-0.23cm} \footnotesize {\#$L$}}
    \psfrag{L-1}{\hspace{-0.3cm} \footnotesize {\#$L\!-\!1$}}
    }
    \vspace{-0.15cm}
    \caption{The time series of end-to-end uplink and downlink data transmissions between $K$ SNs and the AP.}
    \label{Fig:Fig.2}
\end{figure}

It is also assumed that uplink and downlink data transmissions between the SNs and the AP are accomplished within $L$ time slots. In the case of TDD transmission, each slot with the length of $\tau$ is divided into uplink transmission (UT) and downlink transmission (DT) sub-slots with the lengths of $\tau_u$ and $\tau_d\!=\! \tau \!-\! \tau_u $, respectively, c.f. Fig.~\ref{Fig:Fig.2}. For an end-to-end connection between $K$ SNs and the AP, we have the following transmission framework.

\textit{Connection establishment}. To establish the connection, the AUV and UAV relays transmit paging signals to their connected ends and synchronize them through a header slot (HS) with the length of $\tau_\text{HS}$. 

\textit{$L$-slot data transmission}. After the HS, the SNs and AP transmit their uplink and downlink information signals, respectively, in their dedicated sub-slots. Once each relay detects the signals, it {demodulates}, buffers, and {forwards} (DBF) the remodulated signals over the next time slot.

\textit{Connection termination}. Finally, after $L$ time slots, the relays transmit another paging signals in a trailer slot (TS) with the length of $\tau_\text{TS}$ to terminate the connection.

Before analysing the end-to-end performance of the UWSN, in the following subsections, we provide UWOC, FSO, and RF channel models, SNR expressions, and their corresponding statistics.

\subsection{UWOC Channel Model}
The UWOC link of the $k$th SN is modeled by \cite{nabavi2019empirical}
\begin{equation}\label{Eq:Eq1}
    h_k = h_{l,k} h_{t,k} h_{p,k},~\text{for}~k = 1, 2, ..., K,
\end{equation}
which includes the oceanic path-loss, $h_{l,k}$, turbulence, $h_{t,k}$, and pointing error, $h_{p,k}$. 

\subsubsection{Oceanic path-loss} $h_{l,k}$ is modeled under Beer-Lambert law, as below
    \begin{equation}\label{Eq:Eq2}
        h_{l,k} =  \operatorname{exp} \!\left(- {\alpha_{a,k} d_{a,k}} \right)\!,
    \end{equation}
where $\alpha_{a,k}$ denotes the water extinction factor, and $d_{a,k}$ represents the average distance between the $k$th SN and the AUV.

\subsubsection{Oceanic turbulence} For the conventional weak turbulence conditions, $h_{t,k}$ is modeled by a log-normal RV with the following probability density function (p.d.f) \cite{christopoulou2019outage}
\begin{equation}\label{Eq:Eq3}
    f_{h_{t,k}}(h_{t,k}) = \frac{1}{2h_{t,k} \sqrt{2 \pi \sigma_{x_k}^2}} \operatorname{exp} \!\left(\!- \frac{\left(\ln(h_{t,k}) - 2\mu_{x_k}\right)^2}{8 \sigma_{x_k}^2} \right)\!\!,
\end{equation}
such that $\mu_{x_k} \!=\! -\sigma_{x_k}^2$, and $\sigma_{x_k}^2 \!=\! 0.307 C_n^2 k_k^{7/6} d_{a,k}^{11/6}$, where $C_n^2$ represents the index of the refraction structure, $k_k\!=\!{2 \pi}/\lambda_k$ is the optical wave number, and $\lambda_k$ denotes the wavelength of the $k$th WDM channel.

\subsubsection{Oceanic pointing error} The p.d.f of $h_{p,k}$ for a circular detection mechanism is modeled by \cite{naik2020co,sandalidis2008ber}
\begin{equation}\label{Eq:Eq4}
    f_{h_{p,k}}(h_{p,k}) = \frac{\xi^2}{h_0^{\xi^2}} h_{p,k}^{\xi^2-1}, ~~ 0\leq h_{p,k}\leq h_0,
\end{equation}
where $h_0$ and $\xi$ are UWOC pointing error constants.

\subsubsection{Statistics of the UWOC link}
By using (\ref{Eq:Eq2})--(\ref{Eq:Eq4}) and similar steps as \cite{farid2007outage}, the p.d.f of $h_k$ is derived as
\begin{equation}\label{Eq:Eq6}
    f_{h_k}(h_k) = \frac{\xi^2 h_k^{\xi^2-1}}{2 (h_0h_{l,k})^{\xi^2}} \operatorname{erfc}\! \left(\! \frac{\ln\!\left(\!{\frac{h_k}{h_0h_{l,k}}}\!\right) + \varphi_k}{\sqrt{8 \sigma_{x_k}^2}}\!\right) \!\varphi_k^\prime,
\end{equation}
where $\varphi_k \!=\! 2 \sigma_{x_k}^2\! \left(1+2\xi^2\right)$, and $\varphi_k^\prime \!=\! 2 \sigma_{x_k}^2 \xi^2 \! \left(1+\xi^2\right)$. 

Therefore, the received signal transmitted over the $k$th UWOC link is written as
\begin{equation}\label{Eq:Eq7}
    r_k = \eta_{k} h_{k} s_k + n_k,
\end{equation}
where $\eta_k$ denotes the optical-to-electrical conversion parameter, $s_k \!\in\! \{0,\sqrt{P}\}$ is the OOK modulation symbol, $P$ denotes the maximum transmission power, and $n_k \!\sim\! \mathcal{N}(0,\delta_k^2)$ indicates the additive noise. Thus, the SNR of the $k$th UWOC link is derived as
\begin{equation}\label{Eq:Eq8}
    \gamma_k = \frac{\left(\eta_{k} h_{k} s_k\right)^2}{\delta_k^2}
    = \bar{\gamma}_k h_k^2,
\end{equation}
where $\bar{\gamma}_k = {\eta_k^2 s_k^2}/{\delta_k^2}$ represents the average SNR. By the use of (\ref{Eq:Eq6}) and (\ref{Eq:Eq8}), the p.d.f of $\gamma_k$ is obtained as
\begin{equation}\label{Eq:Eq9}
    f_{\gamma_k}(\gamma_k) = \frac{\xi^2 \left(\gamma_k/\bar{\gamma}_k\right)^{\frac{\xi^2}{2}-1}}{4 (h_0h_{l,k})^{\xi^2}} \operatorname{erfc}\! \left(\! \frac{\ln\!\left(\!{\frac{\sqrt{\gamma_k}}{h_0 h_{l,k} \sqrt{\bar{\gamma}_k}}}\!\right) + \varphi_k}{\sqrt{8 \sigma_{x_k}^2}}\!\right) \!\varphi_k^\prime.
\end{equation}
By the use of the standard CDF definition, and after some mathematical manipulations, the CDF of $\gamma_k$ is derived as
\begin{align}\label{Eq:Eq10} \nonumber
    &F_{\gamma_k}(\gamma_k) = \frac{\bar{\gamma}_k \varphi_k^\prime}{2}  \Bigg[\!\operatorname{erfc}\! \left( \frac{\Theta_k(\gamma_k)}{\sqrt{8 \sigma_{x_k}^2}} \right) \! \operatorname{exp}\! \Big({{\Theta_k(\gamma_k)\xi^2}}\Big) \\
    &~~~+
    \operatorname{erfc}\! \left(\frac{4 \sigma_{x_k}^2 \xi^2 - \Theta_k(\gamma_k)}{\sqrt{8 \sigma_{x_k}^2}} \right) \! \operatorname{exp}\! \Big(2 \sigma_{x_k}^2 \xi^4\Big)\!\Bigg]\! \operatorname{exp}\! \Big(\!\!-\varphi_k \xi^2\Big)\!,
\end{align}
where
$
    \Theta_k(\gamma_k) = \ln\! \left(\!{\frac{\sqrt{\gamma_k}}{h_0 h_{l,k} \sqrt{\bar{\gamma}_k}}}\!\right) + \varphi_k.
$

\subsection{FSO Channel Model}
The AUV--UAV FSO links experience W2A and A2W impacts caused by the erratic random and non-random waves in the air-water interface at the uplink and downlink transmissions, respectively. Specifically, the waves reflect and scatter optical signals and result in extra additive loss component which can exceed the absorption loss. Therefore, the FSO channel is modeled as
\begin{equation}\label{Eq:Eq11}
   I_j = I_{l,j} I_{t,j} I_{p,j},
\end{equation}
where $j\!=\!1$ stands for the downlink transmission, while $j\!=\!2$ indicates the uplink one. Also, $I_{l,j}$, $I_{t,j}$, and $I_{p,j}$ denote the atmospheric path-loss, turbulence, and pointing error, respectively.

\subsubsection{Atmospheric path-loss} $I_{l,j}$ is given by
\begin{equation}\label{Eq:Eq12}
    I_{l,j} = \operatorname{exp} \!\left( - \alpha_{au} d_{au} \right)\!,
\end{equation}
where $\alpha_{au}$ is the air attenuation factor which depends on weather conditions, and $d_{au}$ denotes the average distance between the AUV and UAV.

\subsubsection{Atmospheric turbulence} By taking into account the A2W and W2A impacts, $I_{t,j}$ follows the Birnbaum-Saunders distribution with the following p.d.f\cite{nabavi2019empirical,nabavi2019performance}
\begin{align}\label{Eq:Eq13} \nonumber
   f_{I_{t,j}}(I_{t,j};\alpha, \beta) &= \frac{1}{2 \sqrt{2 \pi} \alpha \beta}  \left[\left(\frac{\beta}{I_{t,j}}\right)^{\!\!1/2} \! + \left(\frac{\beta}{I_{t,j}}\right)^{\!\!3/2}\right]\\ &~~~\times \operatorname{exp} \!\left[ -\frac{1}{2 \alpha^2} \left( \frac{I_{t,j}}{\beta} + \frac{\beta}{I_{t,j}} - 2\right) \right]\!,
\end{align}
where $\alpha\!>\!0$ and $\beta\!>\!0$ denote shape and scale parameters, respectively.

\subsubsection{Atmospheric pointing error} For $I_{p,j}$, we have
\begin{equation}\label{Eq:Eq14}
    f_{I_{p,j}}(I_{p,j}) = \frac{\zeta^2}{I_0^{\zeta^2}} I_{p,j}^{\zeta^2-1}, ~~ 0\leq I_{p,j}\leq I_0,
\end{equation}
where $I_0\!=\!\left[\operatorname{erf}(\nu)\right]^2$, and $\zeta\!=\!\frac{1}{2} w_{z_{eq}}\sigma_s^{-1}$ denotes the ratio between the equivalent beam radius and FSO pointing error displacement standard deviation. Also, $w_{z_{eq}}^2 \!=\! w_z^2 \sqrt{0.25 \pi} \operatorname{erf}(\nu) \nu^{-1} \operatorname{exp}(\nu^2)$, and $\nu \!=\! \sqrt{0.5 \pi}\, w_z^{-1} r_{s} $, where $w_z$ denotes FSO beam waist at distance $z$, and $r_s$ implies the alignment-based radial distance at the detector.

In the following two subsections, statistics of the FSO link are separately investigated for the DTS and RRS schemes.
\subsubsection{Statistics of the FSO link for the DTS} In the DTS, the signals are transmitted over two independent links in the uplink and downlink. Thus, the statistical properties of those links are studied in what follows.  
\begin{proposition} The p.d.f of the FSO link for the DTS scheme is given by \label{Pro:Pro1}
\begin{align}\label{Eq:Eq15} \nonumber
    f_{I_j}(I_j;\alpha,\beta) &= 
    \frac{e^{1/{\alpha^2}}}{2\sqrt{\pi}}
    \frac{\zeta^2 I_j^{\zeta^2-1}}{\gamma_{0,j}^{\zeta^2}} \Bigg[\Gamma\!\left(\frac{1}{2} - \zeta^2, \frac{I_j}{\gamma_{0,j}} \right)\\
    &~~~+ 
    \frac{1}{2 \alpha^2}\,\Gamma\!\left(\!- \frac{1}{2} - \zeta^2, \frac{I_j}{\gamma_{0,j}} \right) \!\Bigg],
\end{align}
where $\gamma_{0,j} \!=\!2 \alpha^2 \beta I_0I_{l,j}$.
\end{proposition}
\begin{skproof} See Appendix \ref{Ap:Ap.2}.
\end{skproof}
In the DTS, the received FSO signals in the downlink, i.e., $j\!=\!1$, and uplink, i.e., $j\!=\!2$, are given by
\begin{equation}\label{Eq:Eq16}
    \hat{\boldsymbol{r}}_j  = \mu_{j} I_{j} \hat{\boldsymbol{s}}_j + \boldsymbol{w}_j,
\end{equation}
where $\hat{\boldsymbol{s}}_j \!=\! \left[\hat{s}_{j,1},\hat{s}_{j,2},...,\hat{s}_{j,K}\right]^T$, $\hat{s}_{j,k} \!\in\! \{0,\sqrt{\hat{P}_j}\}$ is the information symbol with OOK modulation, and $\mathbb{E}\!\left\{\hat{s}_{j,k}\hat{s}_{j,k^\prime}\right\}\!=\!0$ for $k \!\neq\! k^\prime$. Also, $\mu_{j}$ indicates the optical-to-electrical conversion parameter, and $\boldsymbol{w}_j \!\sim\! \mathcal{N}(\boldsymbol{0},\Phi_j^2I_{\!K\!\times\!K}\!)$ denotes the independent and identically distributed (i.i.d.) additive noise. Thus, the SNR for the $k$th SN is as follows
\begin{equation}\label{Eq:Eq17}
    \hat{{\gamma}}_{j,k} = \frac{\mu_{j}^2 I_j^2 \hat{{s}}_{j,k}^2}{{\Phi_j^2}}
    = \tilde{{\gamma}}_{j,k} I_j^2,
\end{equation}
where $\mathbb{E}\!\left\{\hat{{s}}_{j,k}{w}_{j,k} \right\}\!=\! {0}$, and $\tilde{{\gamma}}_{j,k} \!=\! \mu_{j}^2 \hat{{s}}_{j,k}^2/\Phi_j^2$. By using (\ref{Eq:Eq15}) and (\ref{Eq:Eq17}), we have
\begin{align}\label{Eq:Eq18} \nonumber
    &f_{\hat{\gamma}_{j,k}}(\hat{\gamma}_{j,k};\alpha,\beta) =
    \frac{e^{1/{\alpha^2}}}{4\sqrt{\pi}}
    \frac{\zeta^2 (\hat{\gamma}_{j,k}/\tilde{\gamma}_{j,k}) ^{\frac{\zeta^2}{2}-1}}{\gamma_{0,j}^{\zeta^2}}\\
    &\!\!\times \!\Bigg[\Gamma\!\left(\frac{1}{2} - \zeta^2, \frac{\sqrt{\hat{\gamma}_{j,k}}}{\gamma_{0,j} \sqrt{\tilde{\gamma}_{j,k}}} \right)+
    \frac{1}{2 \alpha^2}\,\Gamma\!\left(\!- \frac{1}{2} - \zeta^2, \frac{\sqrt{\hat{\gamma}_{j,k}}}{\gamma_{0,j} \sqrt{\tilde{\gamma}_{j,k}}} \right) \!\Bigg].\,
\end{align}
By using the CDF definition and after some mathematical manipulations, the CDF of $\hat{\gamma}_{j,k}$ is derived as
\begin{align}\label{Eq:Eq19}  \nonumber
    &F_{\hat{\gamma}_{j,k}}(\hat{\gamma}_{j,k};\alpha,\beta) =\\ \nonumber
    &~~~~~\frac{2e^{1/{\alpha^2}}}{\sqrt{\pi}}
    \frac{\zeta^2 \tilde{\gamma}_{j,k}^{1 - \frac{\zeta^2}{2}}}{\gamma_{0,j}^{\zeta^2}(\zeta^2+6)} 
    \Bigg[\!\Bigg( \!\Gamma\!\left(\frac{1}{2}-\zeta^2,\frac{\sqrt{{\hat{\gamma}_{j,k}}}}{\gamma_{0,j} \sqrt{\tilde{\gamma}_{j,k}}} \right)\\ \nonumber
    &~~~+
     \frac{1}{2\alpha^2}\Gamma\!\left(\!-\frac{1}{2}-\zeta^2,\frac{\sqrt{{\hat{\gamma}_{j,k}}}}{\gamma_{0,j} \sqrt{\tilde{\gamma}_{j,k}}} \right)\!\Bigg) \hat{\gamma}_{j,k}^{\frac{\zeta^2+6}{8}} + \Big(\gamma_{0,j} \sqrt{\tilde{\gamma}_{j,k}}\Big)^{\!\!\frac{\zeta^2+6}{4}} \\
     &\!\times\!
     \Bigg(\!\!\operatorname{\gamma}\!\left(\!2-\frac{3}{4}\zeta^2,\frac{\sqrt{{\hat{\gamma}_{j,k}}}}{\gamma_{0,j} \sqrt{\tilde{\gamma}_{j,k}}} \right) \!+\! 
     \frac{1}{2\alpha^2}\operatorname{\gamma}\!\left(\!1-\frac{3}{4}\zeta^2,\frac{\sqrt{{\hat{\gamma}_{j,k}}}}{\gamma_{0,j} \sqrt{\tilde{\gamma}_{j,k}}} \right)\!\Bigg)
    \!\Bigg].\,
\end{align}

\subsubsection{Statistics of the FSO link for the RRS}
\begin{figure}[t!]
    \centering
    \pstool[scale=0.74]{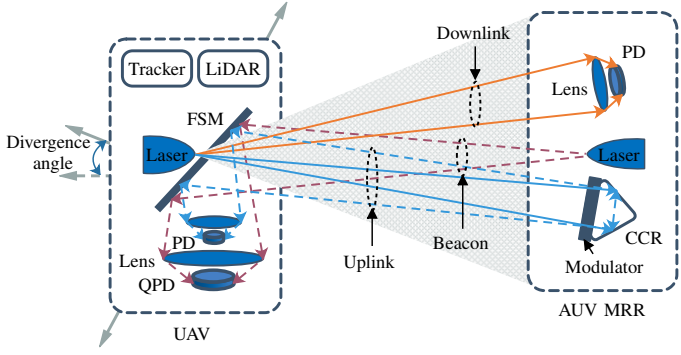}{
    \hspace{-0.5cm}
    \psfrag{Tracker}{\hspace{-0.00cm}\scriptsize Tracker}
    \psfrag{LiDAR}{\hspace{-0.00cm}\scriptsize LiDAR}
    \psfrag{Uplink}{\hspace{-0.00cm}\scriptsize Uplink}
    \psfrag{Downlink}{\hspace{+0.05cm}\scriptsize Downlink}
    \psfrag{Beacon}{\hspace{+0.05cm}\scriptsize Beacon}
    \psfrag{FSM}{\hspace{-0.00cm}\scriptsize FSM}
    \psfrag{Lens}{\hspace{0.00cm}\scriptsize Lens}
    \psfrag{QPD}{\hspace{-0.05cm}\scriptsize QPD}
    \psfrag{PD}{\hspace{+0.05cm}\scriptsize PD}
    \psfrag{Laser}{\hspace{0.00cm}\scriptsize Laser}
    \psfrag{CCR}{\hspace{-0.00cm}\scriptsize CCR}
    \psfrag{Modulator}{\hspace{-0.00cm}\scriptsize Modulator}
    \psfrag{UAV}{\hspace{-0.00cm}\scriptsize UAV}
    \psfrag{AUVMRR}{\hspace{-0.00cm}\scriptsize AUV MRR}
    \psfrag{Gimbal}{\hspace{-0.1cm}\scriptsize Divergence}
    \psfrag{angle}{\hspace{+0.15cm}\scriptsize angle}
    \psfrag{Legend1=Legend1}{\hspace{-0.00cm}\scriptsize}
    }
    \caption{The retro-reflection system among the AUV and UAV relays.}
    \label{Fig:Fig.3}
\end{figure}

Thanks to the MRR system's structure, the pointing error becomes negligible and is not taken into account in the RRS. Hence, for the downlink, by the use of (\ref{Eq:Eq11})--(\ref{Eq:Eq13}) with $j\!=\!1$, we acquire
\begin{align}\label{Eq:Eq21} \nonumber
    f_{I_1}(I_1;\alpha,\beta) &= \frac{1}{2 \sqrt{2 \pi} \alpha \beta I_{l,1}}  \left[\left(\frac{\beta I_{l,1}}{{I_1}}\right)^{\!\!1/2} \! + \left(\frac{\beta I_{l,1}}{{I_1}}\right)^{\!\!3/2}\right] \\
    &~~~\times \operatorname{exp} \!\left[ -\frac{1}{2 \alpha^2} \left( \frac{{I_1}}{\beta I_{l,1}} + \frac{\beta I_{l,1}}{{I_1}} - 2\right) \right]\!.
\end{align}
Consequently, we have
\begin{align}\label{Eq:Eq24} \nonumber
    &f_{\hat{\gamma}_{1,k}}(\hat{\gamma}_{1,k};\alpha,\beta)=\\ \nonumber
    &~~~
    \frac{1}{4 \sqrt{2 \pi} \alpha \beta I_{l,1}}  \left[\left(\frac{\beta I_{l,1} \sqrt{\tilde{{\gamma}}_{1,k}}}{\sqrt{\hat{{\gamma}}_{1,k}}}\right)^{\!\!1/2} \!+ \left(\frac{\beta I_{l,2} \sqrt{\tilde{{\gamma}}_{1,k}}}{\sqrt{\hat{{\gamma}}_{1,k}}}\right)^{\!\!3/2}\right] \\
    &~~~\times \operatorname{exp} \!\left[ -\frac{1}{2 \alpha^2} \left( \frac{\sqrt{\hat{{\gamma}}_{1,k}}}{\beta I_{l,2} \sqrt{\tilde{{\gamma}}_{1,k}}} + \frac{\beta I_{l,2} \sqrt{\tilde{{\gamma}}_{1,k}}}{\sqrt{\hat{{\gamma}}_{1,k}}} - 2\right) \right]\!.
\end{align}
However, for the uplink, the received downlink signal at the AUV is reflected by the CCR and modulated by the MRR terminal, c.f. Fig.~\ref{Fig:Fig.3}.
Therefore, the uplink signal at the UAV's photodiode (PD) is given by\footnote{Conventionally, the equivalent channel model for the RRS scheme is obtained by $I \!=\! I_1I_2$ with correlation coefficient $\rho$. However, we assume that forward and backward channels are reciprocal, i.e., $\rho\!=\!1$, so we have $I \!=\! I_1^2 \!=\! I_2^2$.}
\begin{equation}\label{Eq:Eq20}
    \hat{\boldsymbol{r}}_2  = R \mu_{2} I_2^2 \boldsymbol{x}_m \hat{\boldsymbol{s}}_1 + \boldsymbol{w}_2,
\end{equation}
where $\boldsymbol{x}_m \!= \operatorname{diag} \left[x_{m,1}, x_{m,2}, ..., x_{m,K}\right]$
is the i.i.d. modulated signal matrix at the MRR with $x_{m,k} \!\in\! \{0,1\}$ for the $k$th SN with OOK modulation. Also, $R$ denotes the reflection effect of the CCR\footnote{For the proposed
FSO system, parameter $R$ is equivalent to a geometric loss at the AUV--UAV link, which has the formula as in \cite[(\textcolor{red}{14})]{yang2018performance}.}. Therefore, we have
\begin{align}\label{Eq:Eq22} \nonumber
    f_{I_2}(I_2;\alpha,\beta) &= \frac{I_2^{-1/2}}{4 \sqrt{2 \pi} \alpha \beta I_{l,2}}  \left[\left(\frac{\beta I_{l,2}}{\sqrt{I_2}}\right)^{\!\!1/2} \! + \left(\frac{\beta I_{l,2}}{\sqrt{I_2}}\right)^{\!\!3/2}\right] \\
    &~~~\times \operatorname{exp} \!\left[ -\frac{1}{2 \alpha^2} \left( \frac{\sqrt{I_2}}{\beta I_{l,2}} + \frac{\beta I_{l,2}}{\sqrt{I_2}} - 2\right) \right]\!.~
\end{align}
The SNR's diagonal matrix in the uplink is derived as
\begin{equation}\label{Eq:Eq23}
    \hat{{\gamma}}_{2,k} = \frac{ R^2 \mu_{2}^2 I_2^4 {x}_{m,k}^2  \hat{s}_{2,k}^2}{\Phi_2^2}
    = \tilde{{\gamma}}_{2,k} I_2^4,
\end{equation}
where $\tilde{{\gamma}}_{2,k} = R^2 \mu_{2}^2 {x}_{m,k}^2 \hat{{s}}_{2,k}^2/{\Phi_2^2}$. 
As a result, we obtain
\begin{align}\label{Eq:Eq25} \nonumber
    &\!\!\!f_{\hat{\gamma}_{2,k}}(\hat{\gamma}_{2,k};\alpha,\beta) =\\ \nonumber
    &\!\!\!\frac{\sqrt{\tilde{{\gamma}}_{2,k}}}{8 \sqrt{2 \pi} \alpha \beta I_{l,2} \sqrt{\hat{{\gamma}}_{2,k}}}  \left[\left(\frac{\beta I_{l,2} \sqrt[4]{\tilde{{\gamma}}_{2,k}}}{\sqrt[4]{\hat{{\gamma}}_{2,k}}}\right)^{\!\!1/2} \!\!+ \left(\frac{\beta I_{l,2} \sqrt[4]{\tilde{{\gamma}}_{2,k}}}{\sqrt[4]{\hat{{\gamma}}_{2,k}}}\right)^{\!\!3/2}\right] \\
   &~\!\times \operatorname{exp} \!\left[ -\frac{1}{2 \alpha^2} \left( \frac{\sqrt[4]{\hat{{\gamma}}_{2,k}}}{\beta I_{l,2} \sqrt[4]{\tilde{{\gamma}}_{2,k}}} + \frac{\beta I_{l,2} \sqrt[4]{\tilde{{\gamma}}_{2,k}}}{\sqrt[4]{\hat{{\gamma}}_{2,k}}} - 2\right) \right]\!.~
\end{align}
By using (\ref{Eq:Eq24}) and (\ref{Eq:Eq25}), the CDF of $\hat{\gamma}_{j,k}$ for the RRS scheme is derived as
\begin{align}\label{Eq:Eq26}\nonumber
    &F_{\hat{\gamma}_{j,k}}(\hat{\gamma}_{j,k};\alpha,\beta)=\\
    &~~~
    \operatorname{\Phi}\!\Bigg(\frac{1}{\alpha} \Bigg[ \bigg(\frac{\sqrt[2j]{\hat{{\gamma}}_{j,k}}}{\beta I_{l,j}\sqrt[2j]{\tilde{{\gamma}}_{j,k}}}\bigg)^{\!\!1/2} \!\!- \bigg(\frac{\beta I_{l,j}\sqrt[2j]{\tilde{{\gamma}}_{j,k}}}{\sqrt[2j]{\hat{{\gamma}}_{j,k}}}\bigg)^{\!\!1/2}\Bigg]  \Bigg).
\end{align}
\subsection{RF Channel Model}

The RF link between the UAV and the terrestrial AP is modeled by
\begin{equation}
    G = G_l^{1/2} G_s,
\end{equation}
where $G_l$ and $G_s$ denote the large- and small-scale fading, respectively.

\subsubsection{Large-scale fading} The large-scale fading consists of the pathloss and shadowing, as follows \cite{goldsmith2005wireless}
\begin{equation}
    G_l = -20\log_{10}\!\left(\!\frac{40 \pi}{3}f\!\right) \!-\! 27\log_{10}(d_{ut}) \!+\! \chi_{sh}\, [\text{dB}],
\end{equation}
where $f$ $[\text{GHz}]$ is the RF frequency, $d_{ut}$ $[\text{m}]$ denotes the distance, and $\chi_{sh}\!\sim\!\mathcal{N}(0,\sigma_{sh}^2)$ presents the shadowing.

\subsubsection{Small-scale fading} $G_s$ follows the Nakagami-$m$ distribution with the following p.d.f \cite{anees2019performance}
\begin{equation}\label{Eq:Eq27}
    f_{G_s}(G_s;m) = \frac{2 m^m G_s^{2m-1}}{\Gamma(m) \Omega^m} \operatorname{exp}\!\left(\!- \frac{mG_s^2}{\Omega} \right)\!, 
\end{equation}
where $\Omega \!=\! \mathbb{E}\!\left\{G_s^2\right\}$, and $0.5 \!\leq\! m \!\leq\! \infty$ indicates the Nakagami fading parameter. 

The received RF signal is given by
\begin{equation}\label{Eq:Eq28}
    \doublehat{{\boldsymbol{r}}} = G \doublehat{{\boldsymbol{s}}} + \boldsymbol{v},
\end{equation}
where $\doublehat{{\boldsymbol{s}}} \!=\! \left[\doublehat{{s}}_{1},\doublehat{{s}}_{2},...,\doublehat{{s}}_{K}\right]^T$, $\doublehat{{s}}_{k} \!\in\! \{0,\sqrt{\doublehat{{P}}}\}$ is the transmission symbol with OOK modulation, and $\mathbb{E}\!\left\{\doublehat{{s}}_{k}\doublehat{{s}}_{k^\prime}\right\}\!=\!0$ for $k \!\neq\! k^\prime$. Also, $\boldsymbol{v} \!\sim\! \mathcal{CN}(\boldsymbol{0},\Lambda^2 I_{\!K\!\times\!K}\!)$ is the i.i.d. additive noise. Hence, the SNR is obtained as 
\begin{equation}\label{Eq:Eq29}
    \doublehat{\gamma}_k = \frac{G^2 \doublehat{s}_k^2}{\Lambda^2}
    = \doubletilde{\gamma}_k G^2,
\end{equation}
where $\doubletilde{\gamma}_k \!=\! \doublehat{s}_k^2/\Lambda^2$. By the use of (\ref{Eq:Eq27}) and (\ref{Eq:Eq29}), we have
\begin{equation}\label{Eq:Eq30}
    f_{\hat{\hat{\gamma}}_{k}}(\doublehat{{\gamma}}_k;m) = \frac{m^m {\doublehat{{\gamma}}}_k^{m-1}}{\Gamma(m) \doubletilde{{\gamma}}_k^m G_l^m} \operatorname{exp}\!\left(\!- \frac{m\doublehat{{\gamma}}_{k}}{\doubletilde{{\gamma}}_kG_l}\!\right)\!.
\end{equation}
Consequently, the CDF of $\doublehat{{\gamma}}_{k}$ is given by
\begin{equation}\label{Eq:Eq31}
    F_{\hat{\hat{\gamma}}_{k}}(\doublehat{{\gamma}}_k;m) = \frac{1}{\Gamma(m)} \operatorname{\gamma}\!\left(\!m,\frac{m \doublehat{{\gamma}}_{k}}{\doubletilde{{\gamma}}_k G_l}\!\right)\!.
\end{equation}

\section{AUV--UAV Tracking Procedure and Performance Analyses} \label{Sec:Sec3}
In this section, we firstly propose an AUV--UAV tracking procedure, and then the end-to-end outage probability and average BER expressions are derived.

\subsection{AUV--UAV Tracking Procedure}
To set up reliable and stable FSO communications between the AUV and the UAV, a tracking procedure is required. To this end, we propose Algorithm \ref{Alg:Alg.1}, i.e., an $n$-step acquisition-and-tracking algorithm applied at the UAV, wherein two tracking modes are considered; \emph{coarse} and \emph{fine} tracking modes \cite{kaymak2018survey}. Despite trying to provide LOS light paths between the AUV and the UAV in the coarse tracking mode, non-negligible tracking and pointing errors may remain. The fine tracking mode is also considered to fix this issue and make the tracking more accurate.
\begin{itemize}
    \item In the coarse tracking mode, based on the light detection and ranging (LiDAR) technology, the UAV persistently transmits short but energetic pulses to catches the AUV and track its trajectory by measuring the reflected signals. Afterward, a gimbal tracker at the UAV sweeps the AUV's surface to find its optical lens and align the FSO links between the relays, c.f. Fig.~\ref{Fig:Fig.3}.
    \item In the fine tracking mode, the AUV transmits beacon signals to the UAV after detecting the tracking pulses. Then, at the UAV, the beacon signals are measured by a quadrant photodiode (QPD) and used for driving a fast steering mirror (FSM) under a step track algorithm \cite{nakarach2017comparison}, c.f. Fig.~\ref{Fig:Fig.3}. 
\end{itemize}

\begin{figure}[t!]
    \centering
    \pstool[scale=0.6]{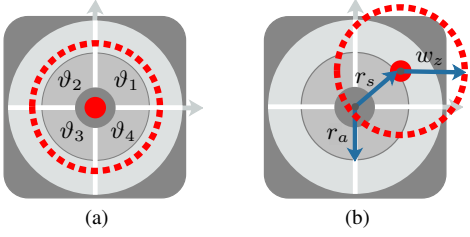}{
    \psfrag{A}{\hspace{-0.00cm}\small $\vartheta_1$}
    \psfrag{B}{\hspace{-0.05cm}\small $\vartheta_2$}
    \psfrag{C}{\hspace{-0.00cm}\small $\vartheta_3$}
    \psfrag{D}{\hspace{-0.05cm}\small $\vartheta_4$}
    \psfrag{Wz}{\hspace{-0.00cm}\small $w_z$}
    \psfrag{rs}{\hspace{-0.1cm}\small $r_s$}
    \psfrag{ra}{\hspace{-0.05cm}\small $r_a$}
    \psfrag{T1}{\hspace{-0.00cm}\footnotesize (a)}
    \psfrag{T2}{\hspace{-0.00cm}\footnotesize (b)}
    \psfrag{T3}{\hspace{-0.00cm}\footnotesize (c)}
    }
    \caption{Four quadratic areas of the QPD, where the bold gray and red dashed circles represent FSO aperture and beam waist, respectively. 
    Herein, (a) and (b) present the perfect alignment, i.e., $(e_x,e_y) \!=\! (0,0)$, and misalignment, i.e., $(e_x,e_y) \!\neq\! (0,0)$, respectively.}
    \label{Fig:Fig.4}
\end{figure}
As depicted in Fig.~\ref{Fig:Fig.4}, the measured FSO beam at four quadrants of the QPD are denoted by $\vartheta_1$, $\vartheta_2$, $\vartheta_3$, and $\vartheta_4$, thus the tracking errors are calculated as \cite{nakarach2017comparison}
\begin{subequations}
\begin{align}\label{Eq:Eq32}
    e_x &= \left| \frac{(\vartheta_1 + \vartheta_4) - (\vartheta_2 + \vartheta_3)}{\vartheta_1 + \vartheta_2 + \vartheta_3 + \vartheta_4} \right|,\\ \label{Eq:Eq33}
    e_y &= \left| \frac{(\vartheta_1 + \vartheta_2) - (\vartheta_3 + \vartheta_4)}{\vartheta_1 + \vartheta_2 + \vartheta_3 + \vartheta_4}\right|.
\end{align}
\end{subequations}
The aim is to achieve almost \emph{perfect} alignment, i.e., $(e_x,e_y) \!\cong\! (0,0)$. 
However, it is more practical to define thresholds, such as $\varepsilon_x$ and $\varepsilon_y$, for indicating alignment quality conditions. Once $e_x$ and $e_y$ individually meet $\varepsilon_x$ and $\varepsilon_y$, the FSO connections between the AUV and the UAV relays will be established and continued.

\begin{algorithm}[t!]
\DontPrintSemicolon
    \caption{\textcolor{black}{The $n$-step tracking procedure}} \label{Alg:Alg.1}
    \KwInput{Angle parameters $\psi_j$, $\psi_\text{min}$, $\psi_\text{max}$, and $\Delta\psi$; the FSO beam at the QPD's quadrants $\vartheta_1$, $\vartheta_2$, $\vartheta_3$, and $\vartheta_4$; QPD sampling size $A_\text{QPD}$; thresholds $\varepsilon_x$ and $\varepsilon_y$; step factors $n_0$ and $n$.}
    Catch the AUV by using the LiDAR technology.\\
    Initiate $i\!=\!0$.\\
    \While{the AUV's beacon signals are not detected}{\nonl\textit{\textcolor{gray}{Coarse tracking:}}\\
    Set the UAV's divergence angle to $\psi_j\!=\!\psi_\text{min}\!+\!i \Delta \psi$.\\
    \lIf{$\psi_j\!=\!\psi_\text{max}$}{go to {\scriptsize{\textbf{2}}};}
     \lElse{$i\!=\!i\!+\!1$.}}
     Set $n_0\!=\!i$.\\
     Initiate $i\!=\!0$.\\
     Compute $(e_x,e_y)$ by $\vartheta_1$, $\vartheta_2$, $\vartheta_3$, $\vartheta_4$, (\ref{Eq:Eq32}), and (\ref{Eq:Eq33}).\\
     \For{$i\!\gets\!0$ \KwTo $A_\text{\normalfont{QPD}}\!-\!1$}{\nonl\textit{\textcolor{gray}{Fine tracking:}}\\
     \lIf{$(e_x,e_y) \!\leq\! (\varepsilon_x,\varepsilon_y)$}{set $n\!=\!i\!+\!n_0$; stop the process;}
     \lElse{adjust the UAV's FSM for the $i$th tracking step; update $(e_x,e_y)$.}}
     Go to {\scriptsize{\textbf{2}}}.
\end{algorithm}

\subsection{Outage Probability}
The end-to-end SNR for the $k$th SN in the downlink, i.e., $j\!=\!1$, and uplink, i.e., $j\!=\!2$, is defined as $\gamma_{j,k} \!=\! \operatorname{min} \!\big\{\!\gamma_k, \hat{\gamma}_{j,k}, \doublehat{{\gamma}}_k \!\big\}$. Herein, $\gamma_k$, $\hat{\gamma}_{j,k}$, and $\doublehat{{\gamma}}_k$ are the SNRs of the UWOC, FSO, and RF links, which are previously presented in (\ref{Eq:Eq8}), (\ref{Eq:Eq17}) or (\ref{Eq:Eq23}), and (\ref{Eq:Eq29}). Therefore, the CDF of $\gamma_{j,k}$ is computed as \cite{papoulis1991random}
\begin{align}\label{Eq:Eq34} \nonumber
    &F_{\gamma_{j,k}}^{}(\gamma_{j,k};\alpha, \beta, m) = F_{\gamma_k}(\gamma_{j,k}) \!+\! F_{\hat{\gamma}_{j,k}}(\gamma_{j,k}; \alpha, \beta)\\ \nonumber
    &~~~~~+\! F_{\hat{\hat{\gamma}}_{k}}(\gamma_{j,k}; m) \!-\! F_{\gamma_k}(\gamma_{j,k}) F_{\hat{\gamma}_{j,k}}(\gamma_{j,k}; \alpha, \beta)\\ \nonumber
    &~~~~~-\! F_{\hat{\gamma}_{j,k}}(\gamma_{j,k}; \alpha, \beta) F_{\hat{\hat{\gamma}}_{k}}(\gamma_{j,k}; m)  \!-\! F_{\gamma_k}(\gamma_{j,k}) F_{\hat{\hat{\gamma}}_{k}}(\gamma_{j,k}; m)\\ \nonumber
    &~~~~~+ F_{\gamma_k}(\gamma_{j,k}) F_{\hat{\gamma}_{j,k}}(\gamma_{j,k}; \alpha, \beta) F_{\hat{\hat{\gamma}}_{k}}(\gamma_{j,k}; m) \\
    &= 1 \!-\! \Big(\!1 \!-\! F_{\gamma_k}(\gamma_{j,k})\!\Big)
    \Big(\!1 \!-\! F_{\hat{\gamma}_{j,k}}(\gamma_{j,k}; \alpha, \beta)\!\Big)
    \Big(\! 1\!-\! F_{\hat{\hat{\gamma}}_{k}}(\gamma_{j,k}; m) \!\Big).
\end{align}
The quality of service (QoS) is ensured by keeping $\gamma_{j,k}$ above a given threshold $\gamma_{th}$. Thus, the outage probability for the $k$th SN is defined as
\begin{eqnarray}\label{Eq:Eq35}
    P_{out,j,k}(\alpha,\beta,m) \triangleq \operatorname{Pr}\! \left\{\gamma_{j,k} \leq \gamma_{th} \right\} = F_{\gamma_{j,k}}^{}(\gamma_{th};\alpha, \beta, m).
\end{eqnarray}
By using (\ref{Eq:Eq34}) and (\ref{Eq:Eq35}), we have
\begin{align}\label{Eq:Eq36} \nonumber
    &P_{out,j,k}(\alpha,\beta,m) = 1 \!-\! \Big(\! 1 \!-\! F_{\gamma_k}(\gamma_{th}) \!\Big) \\
    &~~~~~~~~ \times\! \Big(\! 1 \!-\! F_{\hat{\gamma}_{j,k}}(\gamma_{th}; \alpha, \beta) \!\Big)
      \Big(\! 1 \!-\! F_{\hat{\hat{\gamma}}_{k}}(\gamma_{th}; m)\!\Big).
\end{align}

\subsection{Average Bit Error Rate} 
As well-known, the bit error probability at each hop becomes independent from that of the other link according to the DBF protocol. Hence, the end-to-end average BER for the $k$th SN in the downlink, i.e., $j\!=\!1$, and uplink, i.e., $j\!=\!2$, is derived as
\begin{align}\label{Eq:Eq37} \nonumber
    &P_{e,j,k}(\alpha,\beta,m) =\\
    &~~~~~~ 1\!-\! \Big(\!1 \!-\! P_{e_1,k}\!\Big)\Big(\!1 \!-\! P_{e_2,j,k}(\alpha,\beta)\!\Big)\Big(\!1\!-\! P_{e_3,k}(m)\!\Big),
\end{align}
where $P_{e_1,k}$, $P_{e_2,j,k}(\alpha,\beta)$, and $P_{e_3,k}$ respectively denote the average BERs of the SN--AUV, AUV--UAV, and UAV--AP links, which have the following expressions.
\begin{align}\label{Eq:Eq38} \nonumber
   P_{e_1,k} &= \frac{q^p}{2 \Gamma(p)} \int_{0}^{\infty} e^{-q \gamma} \gamma^{p-1} F_{{\gamma}_{k}}(\gamma)\, d\gamma \\ \nonumber
    &= \frac{q^p\bar{\gamma}_k \varphi_k^\prime}{2 \Gamma(p)}\operatorname{exp}\! \big(\!-\!\varphi_k \xi^2\big) \Bigg[2 \operatorname{exp}\! \Big(-\frac{c_1(8 \sigma_{x_k}^2\xi^2+c_1)}{8 \sigma_{x_k}^2}\Big) \\ \nonumber
    &\times\!
        \Bigg(\dfrac{\operatorname{exp}\! \Big(\big(8 \sigma_{x_k}^2(\xi^2+2p)+2c_1\big)^2/{32 \sigma_{x_k}^2}\Big)}{\xi^2+2p} \\ \nonumber
    &- \dfrac{\operatorname{exp}\! \Big(\big(8 \sigma_{x_k}^2(\xi^2+2p+2)+2c_1\big)^2/{32 \sigma_{x_k}^2}\Big)}{\xi^2+2p+2}\Bigg)\\ \nonumber
    &+  \operatorname{exp}\! \Big(2 \sigma_{x_k}^2 \xi^4\Big)\! \Bigg(\dfrac{\operatorname{exp}\! \Big((p+1)\big(8 \sigma_{x_k}^2(p+1)+2c_2\big)\Big)}{p+1} \\ 
    &- \dfrac{\operatorname{exp}\! \Big(p\big(8 \sigma_{x_k}^2p+2c_2\big)\Big)}{p}\Bigg) \Bigg],
\end{align}
where $c_1\!=\!\ln\! \big({{h_0 h_{l,k} \sqrt{\bar{\gamma}_k}}}\big) \!-\! \varphi_k$, and $c_2\!=\!c_1 \!+ {4 \sigma_{x_k}^2}\xi^2$. 
However, $P_{e_2,j,k}(\alpha,\beta)$ is computed separately for the DTS and RRS schemes. Hence, in the DTS, we have
\begin{align}\label{Eq:Eq39} \nonumber
   P_{e_2,j,k}(\alpha,\beta) &=\frac{q^p}{2 \Gamma(p)}\!\int_{0}^{\infty}\!\! e^{-q \gamma} \gamma^{p-1} F_{\hat{\gamma}_{j,k}}(\gamma; \alpha, \beta)\, d\gamma \\ \nonumber
    &\hspace{-1.2cm}=\frac{q^pe^{1/{\alpha^2}}}{ \sqrt{\pi}\Gamma(p)} \frac{\zeta^2 \tilde{\gamma}_{j,k}({\gamma_{0,j}^2\tilde{\gamma}_{j,k}})^{- \frac{\zeta^2}{2}+p+1}}{(\zeta^2+6)}\Bigg[
        (\gamma_{0,j}^2{\tilde{\gamma}_{j,k}})^{q} \\ \nonumber
    &\hspace{-1.2cm}\times\! \bigg(\!- \frac{\Gamma(-\xi^2+2c_3+\frac{9}{2})}{c_3+2} - \frac{\Gamma(-\xi^2+2c_3+\frac{7}{2})}{2\alpha^2(c_3+2)} \\ \nonumber &\hspace{-1.2cm}+\frac{\Gamma(-\frac{3}{4}\xi^2+2p+4)}{p+1} - \frac{\Gamma(-\frac{3}{4}\xi^2+2p+3)}{2\alpha^2(p+1)} \bigg) \\ \nonumber
    &\hspace{-1.2cm}+ \bigg( \frac{\Gamma(-\xi^2+2c_3+\frac{5}{2})}{c_3+1} + \frac{\Gamma(-\xi^2+2c_3+\frac{3}{2})}{2\alpha^2(c_3+1)} \\  &\hspace{-1.2cm}-\frac{\Gamma(-\frac{3}{4}\xi^2+2p+2)}{p} + \frac{\Gamma(-\frac{3}{4}\xi^2+2p+1)}{2\alpha^2p} \bigg) \Bigg],
\end{align}
where $c_3\!=\!\frac{\zeta^2-2}{8}\!+\!p$, and in the RRS, we have
\begin{align}\label{Eq:Eq40} \nonumber
    &P_{e_2,j,k}(\alpha,\beta)= \frac{q^p}{2 \Gamma(p)}\!\int_{0}^{\infty}\!\! \!\! e^{-q \gamma} \gamma^{p-1} F_{\hat{\gamma}_{j,k}}(\gamma; \alpha, \beta)\, d\gamma \\ \nonumber
    &= \frac{q^p}{4 \Gamma(p)}\!\int_{0}^{\infty}\! e^{-q \gamma} \gamma^{p-1} \operatorname{erfc}\!\Bigg(\!\!\frac{1}{\alpha \sqrt{2}} \Bigg[\! \bigg(\frac{\sqrt[2j]{{{\gamma}}}}{\beta I_{l,j}\sqrt[2j]{\tilde{{\gamma}}_{j,k}}}\bigg)^{\!\!1/2}\\ \nonumber
    &~~~- \bigg(\frac{\beta I_{l,j}\sqrt[2j]{\tilde{{\gamma}}_{j,k}}}{\sqrt[2j]{{{\gamma}}}}\bigg)^{\!\!1/2}\Bigg]  \Bigg) d\gamma \\ \nonumber
    &\simeq \frac{q^p}{4 \Gamma(p)} \Bigg[\! -\left(\!{\alpha\sqrt{2\beta I_{l,j}\sqrt[2j]{\tilde{{\gamma}}_{j,k}}}}\right)^{\!4j(p+1)} \frac{\Gamma\!\left({2j(p+1)+\frac{1}{2}}\right)}{\sqrt{\pi}(p+1)}\\ \nonumber
    &~~~+ \left(\!{\alpha\sqrt{2\beta I_{l,j}\sqrt[2j]{\tilde{{\gamma}}_{j,k}}}}\right)^{\!4jp} \frac{\Gamma\!\left({2jp+\frac{1}{2}}\right)}{\sqrt{\pi}p}\\ \nonumber
    &~~~+ \left(\frac{1}{\alpha} \sqrt{\frac{\sqrt[2j]{\tilde{{\gamma}}_{j,k}}}{2\beta I_{l,j}}}\right)^{\!4j(p+1)} \frac{\Gamma\!\left({-2j(p+1)+\frac{1}{2}}\right)}{\sqrt{\pi}(p+1)}\\
    &~~~- \left(\frac{1}{\alpha} \sqrt{\frac{\sqrt[2j]{\tilde{{\gamma}}_{j,k}}}{2\beta I_{l,j}}}\right)^{\!4jp} \frac{\Gamma\!\left({-2jp+\frac{1}{2}}\right)}{\sqrt{\pi}p} \Bigg].
\end{align}
Also, we have
\begin{align}\label{Eq:Eq41} \nonumber
     P_{e_3,k}(m) &=\frac{q^p}{2 \Gamma(p)} \int_{0}^{\infty} e^{-q \gamma} \gamma^{p-1} F_{\hat{\hat{\gamma}}_{k}}(\gamma; m)\, d\gamma \\ \nonumber
    &= \frac{q^p}{2 \Gamma(p) \Gamma(m)} \int_{0}^{\infty} e^{-q \gamma} \gamma^{p-1}  \operatorname{\gamma}\!\left(\!m,\frac{m \gamma}{\doubletilde{{\gamma}}_k}\!\right) d\gamma \\ \nonumber
    &= \frac{q^p}{2 \Gamma(p) \Gamma(m)} \bigg[ {\left(\frac{\doubletilde{{\gamma}}_k}{m}\right)^{\!p+1}}\frac{\Gamma(m+p+1)}{p+1}\\
    &~~~-  {\left(\frac{\doubletilde{{\gamma}}_k}{m}\right)^{\!p}} \frac{\Gamma(m+p)}{p} \bigg].
    \end{align}
In (\ref{Eq:Eq38})--(\ref{Eq:Eq41}), $p\!=\!0.5$ and $q\!=\!0.25$ with OOK modulation \cite{naik2020co}.


\section{\textcolor{black}{Numerical Results and Discussions}} \label{Sec:Sec4}
\begin{figure}[t!]
    \centering
    \pstool[scale=0.5]{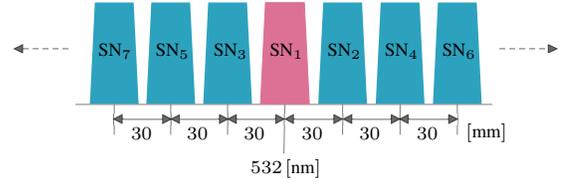}{
    \psfrag{A}{\hspace{-0.08cm}\scriptsize $30$}
    \psfrag{B}{\hspace{-0.38cm}\scriptsize $532$ $\![\text{nm}]$}
    \psfrag{C1}{\hspace{-0.07cm}\scriptsize $\text{SN}_1$}
    \psfrag{C2}{\hspace{-0.08cm}\scriptsize $\text{SN}_2$}
    \psfrag{C3}{\hspace{-0.08cm}\scriptsize $\text{SN}_3$}
    \psfrag{C4}{\hspace{-0.08cm}\scriptsize $\text{SN}_4$}
    \psfrag{C5}{\hspace{-0.08cm}\scriptsize $\text{SN}_5$}
    \psfrag{C6}{\hspace{-0.08cm}\scriptsize $\text{SN}_6$}
    \psfrag{C7}{\hspace{-0.08cm}\scriptsize $\text{SN}_7$}
    \psfrag{ISW=}{\hspace{-0.1cm}\scriptsize $[\text{mm}]$}
    }
    \caption{The WDM diagram for the UWOC links.}
    \label{Fig:Fig.6}
\end{figure}

In what follows, the performance of the triple-hop UWSN is investigated through numerical results, from various perspectives. The wavelength assignments for the UWOC links are performed with the center wavelength of $532$ $\![\text{nm}]$ and grid size of $30$ $\![\text{mm}]$, c.f. Fig.~\ref{Fig:Fig.6}. Thus, the wavelength assigned to the $k$th UWOC link, i.e., $\lambda_k$, is
\begin{equation}\label{Eq:Eq42}
    \dfrac{1}{\lambda_k} = \dfrac{1}{532} + \dfrac{(-1)^{k-1}}{30\!\times\!10^6} \Big\lceil\dfrac{k-1}{2}\Big\rceil.
\end{equation}
The parameters used for the numerical results are summarized in Table~\ref{Tab:Tab.3}, otherwise they are clearly mentioned in the paper.
\noindent
\renewcommand{\arraystretch}{1}
\begin{table}[t!]
\begin{center}
\caption{Network parameters for numerical results.}\label{Tab:Tab.3}
\begin{tabular}{ | l | c | l |}
\hline
\multicolumn{1}{|c|}{\small \text{Parameter}} & \small \text{Symbol} & \multicolumn{1}{c|}{\small \text{Value}}\\
\hline
\hline
\footnotesize FSO wavelengths & \footnotesize $\{\lambda_u,\lambda_d\}$ & \footnotesize  $\{1064,1550\}$ $\![\text{nm}]$\\
\hline
\footnotesize RF frequency & \footnotesize $f$ & \footnotesize  $2$ $\![\text{GHz}]$\\
\hline
\footnotesize Average links' lengths & \footnotesize \!\!\!$\{d_{a,k}, d_{au}, d_{ut}\}$\!\!\!  & \footnotesize $\{0.15, 1.5, 20\}$ $\![\text{km}]\!\!$\\
\hline
\footnotesize Clear water extinction factor& \footnotesize $\alpha_{a,k}$ & \footnotesize $21.79$ $\![\text{dB/km}]$\\
\hline
\footnotesize Clear air attenuation factor& \footnotesize \multirow{3}{*}{$\alpha_{au}$} & \footnotesize $0.44$ $\![\text{dB/km}]$\\
\cline{1-1}\cline{3-3}
\footnotesize Snowy air attenuation factor& \footnotesize & \footnotesize $4.53$ $\![\text{dB/km}]$\\
\cline{1-1}\cline{3-3}
\footnotesize Foggy air attenuation factor& \footnotesize & \footnotesize $50$ $\![\text{dB/km}]$\\
\hline
\footnotesize Refraction structure index& \footnotesize $C_n^2$ & \footnotesize $10^{-15}$ $\![\text{m}^{-2/3}]$ \\
\hline
\footnotesize  UWOC pointing error constants \!\!\! & \footnotesize $\{h_0,\xi\}$ & \footnotesize $\{0.0764,2.35\}$\\
\hline
\footnotesize  FSO displacement deviation & \footnotesize $\sigma_s$  & \footnotesize  $30$\,$[\text{cm}]$\\
\hline
\footnotesize  FSO beam waist & \footnotesize $w_z$  & \footnotesize  $1.25$\,$[\text{m}]$\\
\hline
\footnotesize  FSO receiver's radius & \footnotesize $r_a$  & \footnotesize  $20$\,$[\text{cm}]$\\
\hline
\footnotesize Birnbaum-Saunders parameters& \footnotesize $\{\alpha, \beta\}$ & \footnotesize $\!\{0.6866, 0.8093\}$ \\
\hline
\footnotesize Shadowing standard deviation& \footnotesize $\sigma_{sh}$ & \footnotesize $8$ $\![\text{dB}]$\\
\hline
\footnotesize Nakagami fading parameter & \footnotesize $m$ & \footnotesize $0.5$\\
\hline
\footnotesize  Reflection effect of the CCR\! & \footnotesize $R$ & \footnotesize $0.5$\\
\hline
\end{tabular}
\medskip
\end{center}
\end{table}

\begin{figure}[t!]
\centering
\subfloat[]{
\pstool[scale=0.55]{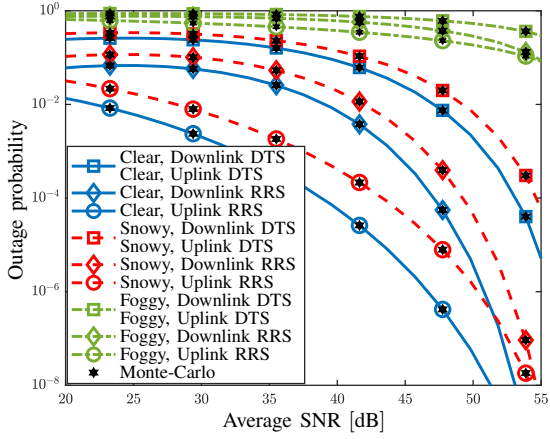}{
\psfrag{OutageProb}{\hspace{-0.52cm}\footnotesize Outage probability}
\psfrag{AverageSNR}{\hspace{-0.45cm}\footnotesize Average SNR $[\text{dB}]$}
\psfrag{DTS-UClear123456478910111}{\scriptsize \text{Clear, Downlink DTS}}
\psfrag{DTS-DClear}{\scriptsize \text{Clear, Uplink DTS}}
\psfrag{RRS-UClear}{\scriptsize  \text{Clear, Downlink RRS}}
\psfrag{RRS-DClear}{\scriptsize \text{Clear, Uplink RRS}}
\psfrag{DTS-USnowy}{\scriptsize \text{Snowy, Downlink DTS}}
\psfrag{DTS-DSnowy}{\scriptsize \text{Snowy, Uplink DTS}}
\psfrag{RRS-USnowy}{\scriptsize  \text{Snowy, Downlink RRS}}
\psfrag{RRS-DSnowy}{\scriptsize \text{Snowy, Uplink RRS}}
\psfrag{DTS-UFoggy}{\scriptsize \text{Foggy, Downlink DTS}}
\psfrag{DTS-DFoggy}{\scriptsize \text{Foggy, Uplink DTS}}
\psfrag{RRS-UFoggy}{\scriptsize  \text{Foggy, Downlink RRS}}
\psfrag{RRS-DFoggy}{\scriptsize \text{Foggy, Uplink RRS}}
\psfrag{Monte-Carlo}{\scriptsize \text{Monte-Carlo}}
}}
\hfill
\subfloat[]{
\pstool[scale=0.55]{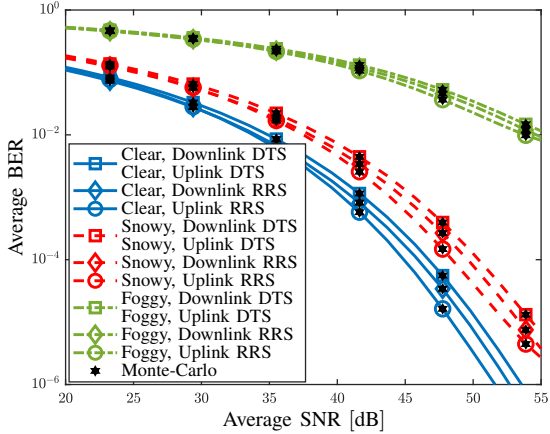}{
\psfrag{ABER}{\hspace{-0.5cm}\footnotesize Average BER}
\psfrag{AverageSNR}{\hspace{-0.45cm}\footnotesize Average SNR $[\text{dB}]$}
\psfrag{DTS-UClear123456478910111}{\scriptsize \text{Clear, Downlink DTS}}
\psfrag{DTS-DClear}{\scriptsize \text{Clear, Uplink DTS}}
\psfrag{RRS-UClear}{\scriptsize  \text{Clear, Downlink RRS}}
\psfrag{RRS-DClear}{\scriptsize \text{Clear, Uplink RRS}}
\psfrag{DTS-USnowy}{\scriptsize \text{Snowy, Downlink DTS}}
\psfrag{DTS-DSnowy}{\scriptsize \text{Snowy, Uplink DTS}}
\psfrag{RRS-USnowy}{\scriptsize  \text{Snowy, Downlink RRS}}
\psfrag{RRS-DSnowy}{\scriptsize \text{Snowy, Uplink RRS}}
\psfrag{DTS-UFoggy}{\scriptsize \text{Foggy, Downlink DTS}}
\psfrag{DTS-DFoggy}{\scriptsize \text{Foggy, Uplink DTS}}
\psfrag{RRS-UFoggy}{\scriptsize  \text{Foggy, Downlink RRS}}
\psfrag{RRS-DFoggy}{\scriptsize \text{Foggy, Uplink RRS}}
\psfrag{Monte-Carlo}{\scriptsize \text{Monte-Carlo}}
}}
\caption{The network's end-to-end (a) outage probability and (b) average BER versus average SNR under different weather conditions.}
\label{Fig:Fig.7}
\end{figure}
To investigate the network's end-to-end outage probability and average BER, Fig.~\ref{Fig:Fig.7}\,(a) and Fig.~\ref{Fig:Fig.7}\,(b) are depicted, respectively. It is shown that the retro-reflection system improves the network's outage probability and average BER, compared to the FSO direct transmission scheme. For example, the RRS scheme outperforms the DTS one in the uplink and downlink with on average $200\%$ ($32\%$) and $80\%$ ($17\%$) better outage probability (average BER), respectively, in the case of the clear weather and the average SNR of $40$ $\![\text{dB}]$. This happens due to the fact that the pointing error is assumed negligible for the FSO links based on the retro-reflection system. The same reason is also true if we compare the RRS uplink with its downlink, where the primary one surpasses the latter one. As an illustrative example, for the clear weather and the average SNR of $40$ $\![\text{dB}]$, the RRS uplink offers about $67\%$ ($13\%$) lower outage probability (average BER) than the downlink one. In this case, since the uplink channel incorporates two correlated pointing error-free forward and backward channels, it presents a better performance compared to the downlink channel. Moreover, in both figures, the network's performance is analyzed under clear, snowy, and foggy weather conditions. As expected, the network's performance degrades by changing the weather from a clear condition to the foggy one. All results are verified by Monte-Carlo simulations with $500$ iterations.

\begin{figure}[t!]
\centering
\pstool[scale=0.55]{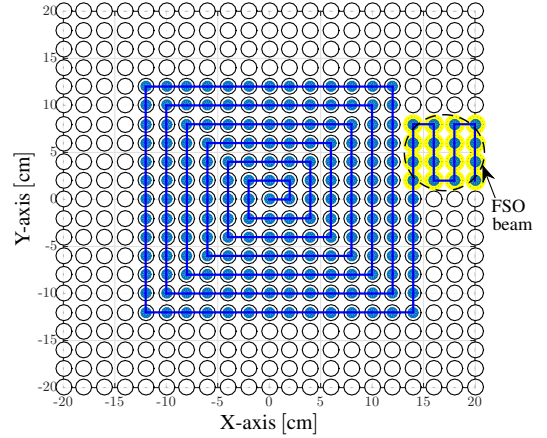}{
\psfrag{X-axis}{\hspace{-0.35cm}\footnotesize X-axis $\![\text{cm}]$}
\psfrag{Y-axis}{\hspace{-0.32cm}\footnotesize Y-axis $\![\text{cm}]$}
\psfrag{FSO}{\hspace{0.05cm}\scriptsize FSO}
\psfrag{beam}{\hspace{0.15cm}\scriptsize beam}}
\caption{The fine tracking procedure on a QPD with a $40\!\times\!40$ $\![\text{cm}^2]$ dimension, $2$ $\![\text{cm}]$ grid size, and $A_\text{\normalfont{QPD}}\!=\!441$ sampling points, i.e., the white spots. Herein, the $16$-point dashed circle area shows a FSO beam refracted on the QPD's 4-quadrant surface, and solid blue spots indicate the tracking points.}
\label{Fig:Fig.8}
\end{figure}

Fig.~\ref{Fig:Fig.8} illustrates the AUV--UAV fine tracking procedure on a QPD with a $40\!\times\!40$ $\![\text{cm}^2]$ dimension, $2$ $\![\text{cm}]$ grid size, and $A_\text{\normalfont{QPD}}\!=\!441$ sampling points. In this figure, a random $16$-point dashed circle area within the defined dimension is assigned for the FSO beam refracted on the QPD after the coarse tracking procedure. Based on the step track algorithm, a tracking pointer starts from the center and sweeps the QPD's surface step-by-step to find the location of the beam. Once the beam is tracked, the pointer tries to measure the beams' territory. After the tracking procedure is successfully done, the FSM aligns the FSO beam to the center of the QPD and minimizes the tracking and pointing errors. Herein, the white, yellow (within the dashed circle area), and solid blue spots show sampling points, FSO beam, and tracking points, where each blue line connecting a tracking point to its neighbor is a tracking step.

\begin{figure}[t!]
\centering
\pstool[scale=0.55]{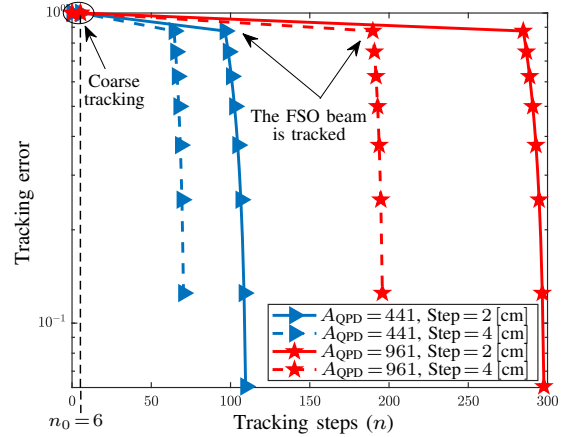}{
\psfrag{Trackingerror}{\hspace{-0.23cm}\footnotesize Tracking error}
\psfrag{Iterationsize}{\hspace{-0.46cm}\footnotesize Tracking steps ($n$)}
\psfrag{A1234567891011121314151617181}{\scriptsize $A_\text{\normalfont{QPD}}\!=\!441$, $\text{Step}\!=\!2$ $\![\text{cm}]$}
\psfrag{B}{\scriptsize $A_\text{\normalfont{QPD}}\!=\!441$, $\text{Step}\!=\!4$ $\![\text{cm}]$}
\psfrag{C}{\scriptsize $A_\text{\normalfont{QPD}}\!=\!961$, $\text{Step}\!=\!2$ $\![\text{cm}]$}
\psfrag{D}{\scriptsize $A_\text{\normalfont{QPD}}\!=\!961$, $\text{Step}\!=\!4$ $\![\text{cm}]$}
\psfrag{Coarse}{\scriptsize Coarse}
\psfrag{Tracking}{\hspace{-0.05cm}\scriptsize tracking}
\psfrag{FSObeam}{\hspace{-0.47cm}\scriptsize The FSO beam}
\psfrag{iscaught}{\hspace{-0.175cm}\scriptsize is tracked}
\psfrag{n}{\hspace{-0.4cm}\scriptsize $n_0\!=\!6$}
}
\caption{The AUV--UAV FSO tracking error versus the number of tracking steps defined in Algorithm \ref{Alg:Alg.1}.}
\label{Fig:Fig.9}
\end{figure}
Besides, Fig.~\ref{Fig:Fig.9} depicts the AUV-UAV tracking error, i.e., $e_x$ or $e_y$, versus the number of tracking steps, based on a Monte-Carlo simulation with $500$ iterations. For the coarse tracking, we have $\psi_\text{min}\!=\!-\tan^{\!-1}(12.5/d_{au} [\text{m}])$, $\psi_\text{max}\!=\!\tan^{\!-1}(12.5/d_{au} [\text{m}])$, and $\Delta\psi\!=\!2\tan^{\!-1}(w_z/d_{au} [\text{m}])$. Given that, if it is assumed that the AUV's length is about $25$ $\![\text{m}]$, the coarse tracking step would take an integer random value between $1$ and $10$. In this figure, the coarse tracking is performed within almost $n_0\!=\!6$ steps. However, for the fine tracking, the procedure as depicted in Fig.~\ref{Fig:Fig.8} is simulated separately for $40\!\times\!40$ and $60\!\times\!60$ $\![\text{cm}^2]$ QPD dimensions, i.e., $A_\text{\normalfont{QPD}}\!=\!441$ and $961$ with the grid size of $2$ $\![\text{cm}]$, respectively, and the step sizes of $2$ and $4$ $\![\text{cm}]$. It is concluded that there is a trade-off between the tracking error threshold and the number of tracking steps. The number of tracking steps reduces for larger step size, with the penalty of non-zero tracking error threshold. For instance, with $441$ sampling points and $2$ $\![\text{cm}]$ step size, the tracking error drops to zero with average $n\!=\!112$ steps. Nevertheless, with the same sampling points but $4$ $\![\text{cm}]$ step size, the tracking error diminishes to its minimum value of $0.11$ with average $n\!=\!67$ steps. Furthermore, it is shown that increasing the number of sampling points boosts the number of tracking steps, thus it takes much time to track the FSO beam on the QPD.

\begin{figure}[t!]
\centering
\subfloat[]{
\pstool[scale=0.55]{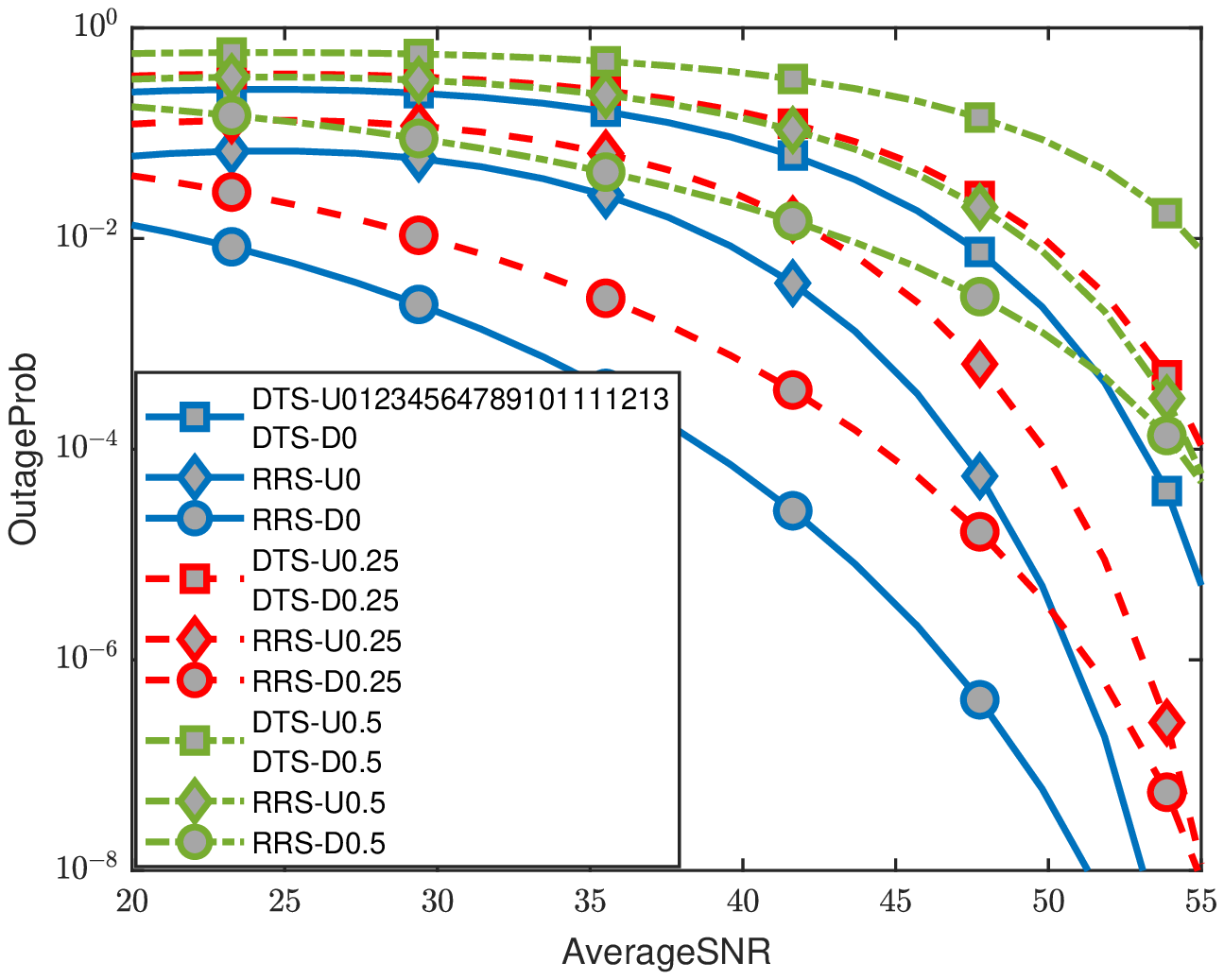}{
\psfrag{OutageProb}{\hspace{-0.52cm}\footnotesize Outage probability}
\psfrag{AverageSNR}{\hspace{-0.45cm}\footnotesize Average SNR $[\text{dB}]$}
\psfrag{DTS-U01234564789101111213}{\scriptsize \text{$\varepsilon\!=\!0$, Downlink DTS}}
\psfrag{DTS-D0}{\scriptsize \text{$\varepsilon\!=\!0$, Uplink DTS}}
\psfrag{RRS-U0}{\scriptsize  \text{$\varepsilon\!=\!0$, Downlink RRS}}
\psfrag{RRS-D0}{\scriptsize \text{$\varepsilon\!=\!0$, Uplink RRS}}
\psfrag{DTS-U0.25}{\scriptsize \text{$\varepsilon\!=\!0.1$, Downlink DTS}}
\psfrag{DTS-D0.25}{\scriptsize \text{$\varepsilon\!=\!0.1$, Uplink DTS}}
\psfrag{RRS-U0.25}{\scriptsize  \text{$\varepsilon\!=\!0.1$, Downlink RRS}}
\psfrag{RRS-D0.25}{\scriptsize \text{$\varepsilon\!=\!0.1$, Uplink RRS}}
\psfrag{DTS-U0.5}{\scriptsize \text{$\varepsilon\!=\!0.2$ Downlink DTS}}
\psfrag{DTS-D0.5}{\scriptsize \text{$\varepsilon\!=\!0.2$, Uplink DTS}}
\psfrag{RRS-U0.5}{\scriptsize  \text{$\varepsilon\!=\!0.2$, Downlink RRS}}
\psfrag{RRS-D0.5}{\scriptsize \text{$\varepsilon\!=\!0.2$, Uplink RRS}}
\psfrag{AA}{\scriptsize $10^0$}
}}
\hfill
\subfloat[]{
\pstool[scale=0.55]{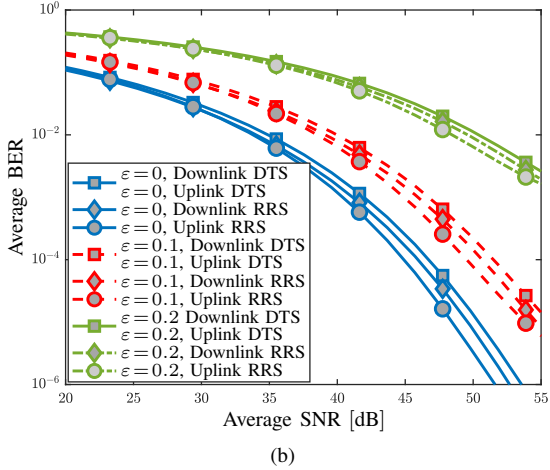}{
\psfrag{ABER}{\hspace{-0.5cm}\footnotesize Average BER}
\psfrag{AverageSNR}{\hspace{-0.45cm}\footnotesize Average SNR $[\text{dB}]$}
\psfrag{DTS-U01234564789101111213}{\scriptsize \text{$\varepsilon\!=\!0$, Downlink DTS}}
\psfrag{DTS-D0}{\scriptsize \text{$\varepsilon\!=\!0$, Uplink DTS}}
\psfrag{RRS-U0}{\scriptsize  \text{$\varepsilon\!=\!0$, Downlink RRS}}
\psfrag{RRS-D0}{\scriptsize \text{$\varepsilon\!=\!0$, Uplink RRS}}
\psfrag{DTS-U0.25}{\scriptsize \text{$\varepsilon\!=\!0.1$, Downlink DTS}}
\psfrag{DTS-D0.25}{\scriptsize \text{$\varepsilon\!=\!0.1$, Uplink DTS}}
\psfrag{RRS-U0.25}{\scriptsize  \text{$\varepsilon\!=\!0.1$, Downlink RRS}}
\psfrag{RRS-D0.25}{\scriptsize \text{$\varepsilon\!=\!0.1$, Uplink RRS}}
\psfrag{DTS-U0.5}{\scriptsize \text{$\varepsilon\!=\!0.2$ Downlink DTS}}
\psfrag{DTS-D0.5}{\scriptsize \text{$\varepsilon\!=\!0.2$, Uplink DTS}}
\psfrag{RRS-U0.5}{\scriptsize  \text{$\varepsilon\!=\!0.2$, Downlink RRS}}
\psfrag{RRS-D0.5}{\scriptsize \text{$\varepsilon\!=\!0.2$, Uplink RRS}}
}}
\caption{The network's end-to-end (a) outage probability and (b) average BER versus average SNR for various tracking error thresholds. Herein, the weather state is assumed to be clear.}
\label{Fig:Fig.10}
\end{figure}
Fig.~\ref{Fig:Fig.10}\,(a) and Fig.~\ref{Fig:Fig.10}\,(b) sequentially present the network's outage probability and average BER versus average SNR for different values of the tracking error threshold, i.e., $\varepsilon\!=\!\varepsilon_x\!=\!\varepsilon_y$. According to Fig.~\ref{Fig:Fig.9}, a minimum number of tracking steps is required based on the number of sampling points and the step size to meet the dedicated threshold. As an example, with $441$ sampling points and $2$ $[\text{cm}]$ step size, about $112$, $109$, and $100$ steps are required to satisfy $\varepsilon\!=\!0$, $\varepsilon\!=\!0.1$, and $\varepsilon\!=\!0.2$, respectively. It is also observed that the RRS uplink outperforms the RRS downlink, and the RRS downlink outperforms both DTS uplink and downlink. It is shown that, by increasing the tracking error threshold, the network's performance degrades since the tracking and pointing errors are enlarged. For instance, in the case of the clear weather with the average SNR of $40$ $\![\text{dB}]$, the outage probability for the uplink RRS takes on average $20\%$ and $75\%$ higher values by increasing the threshold from $\varepsilon\!=\!0$ to $\varepsilon\!=\!0.1$ and $\varepsilon\!=\!0.2$, respectively. For depicting this figure, we initiate $w_z^{-1} r_s$ as $0.84$, $0.88$, and $0.91$ for $\varepsilon\!=\!0$, $\varepsilon\!=\!0.1$, and $\varepsilon\!=\!0.2$, respectively\footnote{One can show that the bounds for the $r_s$ between the \emph{ideal} alignment and \emph{misalignment} are obtained as $1\!-\!w_z^{-1}r_a\!\leq\!w_z^{-1}r_s\!\leq\!1\!+\!w_z^{-1}r_a$ \cite{agheli2021cognitive}.}.

\begin{figure}[t!]
\centering
\subfloat[]{
\pstool[scale=0.55]{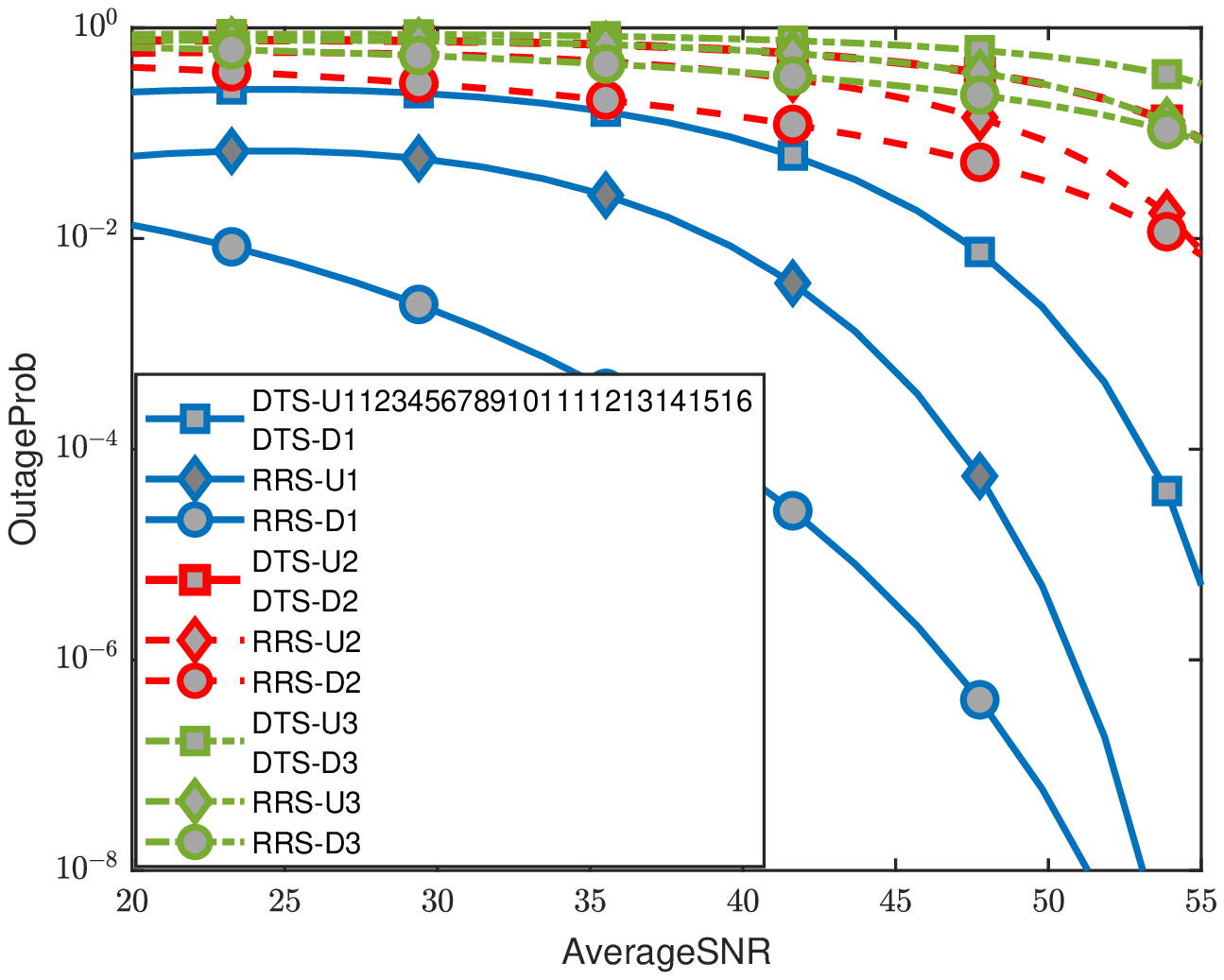}{
\psfrag{OutageProb}{\hspace{-0.52cm}\footnotesize Outage probability}
\psfrag{AverageSNR}{\hspace{-0.45cm}\footnotesize Average SNR $[\text{dB}]$}
\psfrag{DTS-U1123456789101111213141516}{\scriptsize \text{$(1.5,0.15)$, Downlink DTS}}
\psfrag{DTS-D1}{\scriptsize \text{$(1.5,0.15)$, Uplink DTS}}
\psfrag{RRS-U1}{\scriptsize  \text{$(1.5,0.15)$, Downlink RRS}}
\psfrag{RRS-D1}{\scriptsize \text{$(1.5,0.15)$, Uplink RRS}}
\psfrag{DTS-U2}{\scriptsize \text{$(1.5,0.45)$, Downlink DTS}}
\psfrag{DTS-D2}{\scriptsize \text{$(1.5,0.45)$, Uplink DTS}}
\psfrag{RRS-U2}{\scriptsize  \text{$(1.5,0.45)$, Downlink RRS}}
\psfrag{RRS-D2}{\scriptsize \text{$(1.5,0.45)$, Uplink RRS}}
\psfrag{DTS-U3}{\scriptsize \text{$(4.5,0.45)$, Downlink DTS}}
\psfrag{DTS-D3}{\scriptsize \text{$(4.5,0.45)$, Uplink DTS}}
\psfrag{RRS-U3}{\scriptsize  \text{$(4.5,0.45)$, Downlink RRS}}
\psfrag{RRS-D3}{\scriptsize \text{$(4.5,0.45)$, Uplink RRS}}
}}
\hfill
\subfloat[]{
\pstool[scale=0.55]{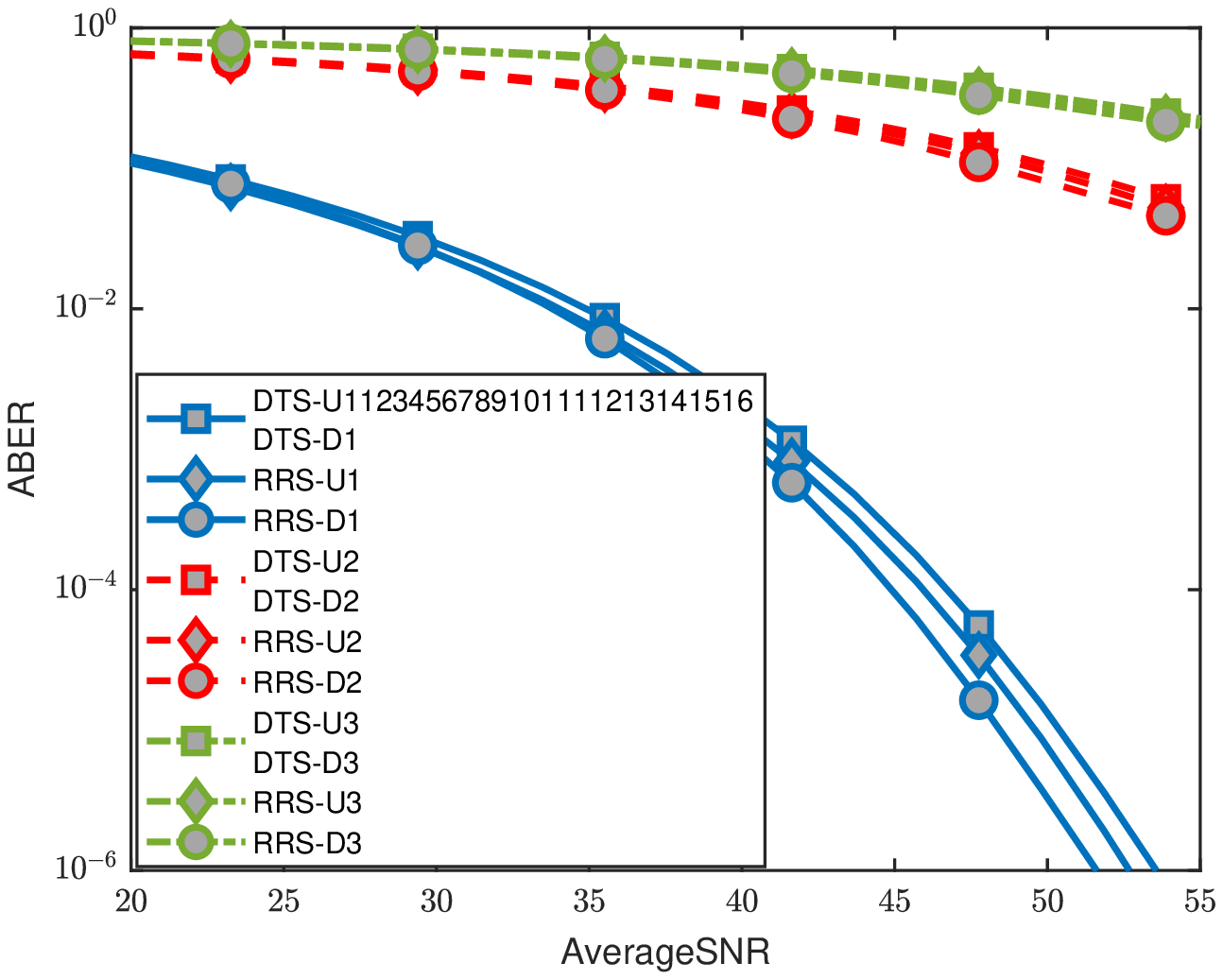}{
\psfrag{ABER}{\hspace{-0.5cm}\footnotesize Average BER}
\psfrag{AverageSNR}{\hspace{-0.45cm}\footnotesize Average SNR $[\text{dB}]$}
\psfrag{DTS-U1123456789101111213141516}{\scriptsize \text{$(1.5,0.15)$, Downlink DTS}}
\psfrag{DTS-D1}{\scriptsize \text{$(1.5,0.15)$, Uplink DTS}}
\psfrag{RRS-U1}{\scriptsize  \text{$(1.5,0.15)$, Downlink RRS}}
\psfrag{RRS-D1}{\scriptsize \text{$(1.5,0.15)$, Uplink RRS}}
\psfrag{DTS-U2}{\scriptsize \text{$(1.5,0.45)$, Downlink DTS}}
\psfrag{DTS-D2}{\scriptsize \text{$(1.5,0.45)$, Uplink DTS}}
\psfrag{RRS-U2}{\scriptsize  \text{$(1.5,0.45)$, Downlink RRS}}
\psfrag{RRS-D2}{\scriptsize \text{$(1.5,0.45)$, Uplink RRS}}
\psfrag{DTS-U3}{\scriptsize \text{$(4.5,0.45)$, Downlink DTS}}
\psfrag{DTS-D3}{\scriptsize \text{$(4.5,0.45)$, Uplink DTS}}
\psfrag{RRS-U3}{\scriptsize  \text{$(4.5,0.45)$, Downlink RRS}}
\psfrag{RRS-D3}{\scriptsize \text{$(4.5,0.45)$, Uplink RRS}}
}}
\caption{The network's end-to-end (a) outage probability and (b) average BER versus average SNR for different UWOC and FSO links' lengths. Herein, the weather state is assumed to be clear.}
\label{Fig:Fig.11}
\end{figure}
Finally, to illustrate the effects of UWOC and FSO links' lengths on the network's performance, Fig.~\ref{Fig:Fig.11} is depicted. For this purpose, we define a set of $(d_{au},d_{a,k})$ and analyze the network's outage probability and average BER for the uplink and downlink transmissions for different values of $d_{au}$ and $d_{a,k}$. Similarly, the RRS scheme enhances the network's performance in comparison to the DTS one in the uplink and downlink. Also, the network's end-to-end outage probability and average BER reduce by increasing the distances of either the UWOC links or the FSO ones due to rising their path-loss coefficients. As an illustrative example, for the average SNR of $40$ $\![\text{dB}]$, the outage probability for the uplink RRS increases on average $198\%$ and $32\%$ by doubling the UWOC and FSO link's lengths, respectively. It goes without saying that changing the distances of the UWOC links affects the network's performance much more than that of the FSO ones, due to the higher water extinction factor compared to the clear weather state.

\section{Conclusion} \label{Sec:Sec5}
We studied the triple-hop UWSN wherein $K$ SNs are connected to the AUV by UWOC links, the AUV is connected to the UAV via FSO links, and the UAV is connected to the terrestrial AP with an RF link. The end-to-end transmission framework was discussed, and the DTS and RRS schemes were considered for the FSO uplink and downlink transmissions, subject to the W2A and A2W impacts. We firstly provided the channel models and their corresponding statistics, then computed the UWSN's end-to-end outage probability and average BER. Furthermore, the AUV--UAV tracking procedure was proposed based on the suggested $n$-step acquisition-and-tracking algorithm with coarse and fine tracking modes, to provide reliable and stable FSO communications. Through numerical results, it was shown that the RSS scheme outperforms the DTS one with on average $200\%$ ($32\%$) and $80\%$ ($17\%$) lower outage probability (average BER) in the uplink and downlink, respectively. Besides, it was concluded that the tracking procedure improves the network's performance with up to $480\%$ and $170\%$, on average, improvements from the outage probability and average BER perspectives, respectively, in comparison to poorly aligned FSO conditions. The results were validated by using Monte-Carlo simulations.
\appendices


\section{The Proof of Proposition \ref{Pro:Pro1}}\label{Ap:Ap.2}
With similar steps as in \cite{farid2007outage}, and by using (\ref{Eq:Eq13}), we have
\begin{align}\label{AP:Eq.2} \nonumber
    &f_{I_j}(I_j;\alpha,\beta) = 
    \frac{\zeta^2 I_j^{\zeta^2-1}}{(I_0I_{l,j})^{\zeta^2}} \int_{I_j/{I_0I_{l,j}}}^{\infty} I_{t,j}^{-\zeta^2} f_{I_{t,j}}(I_{t,j};\alpha,\beta)\, dI_{t,j}\\
    &= \nonumber
    \frac{1}{2 \sqrt{2 \pi} \alpha \beta}
    \frac{\zeta^2 I_j^{\zeta^2-1}}{(I_0I_{l,j})^{\zeta^2}}\\ \nonumber 
    &\times\! \Bigg\{\!\int_{I_j/{I_0I_{l,j}}}^{\infty} \!\!\!\! I_{t,j}^{-\zeta^2}  \left(\frac{\beta}{I_{t,j}}\right)^{\!\!1/2}\! \!\!\operatorname{exp} \!\left[ -\frac{1}{2 \alpha^2} \left( \frac{I_{t,j}}{\beta} + \frac{\beta}{I_{t,j}} - 2\right) \right] \!dI_{t,j} \\ \nonumber
    &+\!
    \int_{I_j/{I_0I_{l,j}}}^{\infty} \!\!\!\! I_{t,j}^{-\zeta^2}  \left(\frac{\beta}{I_{t,j}}\right)^{\!\!3/2} \! \!\!\operatorname{exp} \!\left[ -\frac{1}{2 \alpha^2} \left( \frac{I_{t,j}}{\beta} + \frac{\beta}{I_{t,j}} - 2\right) \right] \!dI_{t,j}\! \Bigg\}\\
    &= \nonumber
    \frac{1}{2 \sqrt{2 \pi} \alpha}
    \frac{\zeta^2 I_j^{\zeta^2-1}}{(\beta I_0I_{l,j})^{\zeta^2}} \\ \nonumber
    &\times\! \Bigg\{\!\int_{I/{\beta I_0I_{l,j}}}^{\infty} \!\! \left(\frac{1}{u}\right)^{\!\!1/2 + \zeta^2}\!\!\! \operatorname{exp} \!\left[ -\frac{1}{2 \alpha^2} \left( u + \frac{1}{u} - 2\right) \right] \!du \\
    &+\!
    \int_{I_j/{\beta I_0I_{l,j}}}^{\infty} \!\! \left(\frac{1}{u}\right)^{\!\!3/2 + \zeta^2}\!\!\! \operatorname{exp} \!\left[ -\frac{1}{2 \alpha^2} \left( u + \frac{1}{u} - 2\right) \right] \!du \!\Bigg\}.
\end{align}
According to the integral limits, the $u$ is large enough to use the approximation $u \!+\! \frac{1}{u} \!\simeq\! u$, at the cost of a negligible error. After some mathematical computations, (\ref{Eq:Eq15}) is derived.

\balance
\bibliographystyle{IEEEtran}
\bibliography{References.bib}

\end{document}